\documentclass{IEEEtran}

\usepackage{amsmath}
\usepackage[utf8]{inputenc}
\usepackage{stmaryrd}
\allowdisplaybreaks
\usepackage{amssymb}
\usepackage{graphicx}
\usepackage{todonotes}
\usepackage{algorithmic}
\usepackage{algorithm}
\usepackage{mathtools}
\usepackage{caption}
\usepackage{subcaption}
\usepackage{float}
\usepackage{comment}
\usepackage{makecell}
\usepackage{xcolor}
\usepackage{multirow}
\usepackage{bm}
\usepackage{bbm}
\usepackage{glossaries}
\newcommand{\vect}[1]{\boldsymbol{\mathrm{#1}}}
\newcommand{\mat}[1]{\boldsymbol{\mathrm{#1}}}

\newcommand{\tr}{\mathrm{tr}}

\newcommand{\diag}{\mathrm{diag}}
\newcommand{\bdiag}{\mathrm{bdiag}}
\newcommand{\rank}{\mathrm{rank}}

\newcommand{\vecop}{\text{vec}}

\newcommand{\norm}[1]{\lVert#1\rVert}
\newcommand{\abs}[1]{\lvert#1\rvert}

\newcommand{\cn}[2]{\ensuremath{\mathcal{C}\cN\left(#1,#2\right)}}

\newcommand{\prob}[1][]{%requires xifthen package%
\ifthenelse{\isempty{#1}}%
      {\ensuremath{P}}%
    {\ensuremath{P\left\(#1\right\)}}%
}

\DeclareMathOperator*{\argmin}{argmin}

\def\defeq{\triangleq}

\newcommand{\transp}{{\sf T}}
\newcommand{\herm}{{\sf H}}

\def\ba{{\vect{a}}}

\def\bee{{\vect{e}}}

\def\bg{{\vect{g}}}
\def\bh{{\vect{h}}}

\def\bu{{\vect{u}}}
\def\bv{{\vect{v}}}
\def\bw{{\vect{w}}}
\def\bx{{\vect{x}}}
\def\by{{\vect{y}}}
\def\bz{{\vect{z}}}
\def\b0{{\vect{0}}}

\def\bzero{{\vect{0}}}
\def\bone{{\vect{1}}}

\def\bphi{{\vect{\phi}}}
\def\bpsi{{\vect{\psi}}}
\def\btheta{{\vect{\theta}}}

\def\bgamma{\vect{\gamma}}
\def\bmu{\vect{\mu}}
\def\blambda{\vect{\lambda}}

\def\bA{{\mat{A}}}

\def\bC{{\mat{C}}}
\def\bD{{\mat{D}}}

\def\bG{{\mat{G}}}

\def\bI{{\mat{I}}}

\def\bR{{\mat{R}}}
\def\bS{{\mat{S}}}

\def\bU{{\mat{U}}}
\def\bV{{\mat{V}}}
\def\bW{{\mat{W}}}
\def\bX{{\mat{X}}}
\def\bY{{\mat{Y}}}

\def\bPsi{{\mat{\Psi}}}
\def\bPhi{{\mat{\Phi}}}

\def\bSigma{\mat{\Sigma}}

\def\bUpsilon{\mat{\Upsilon}}

\def\cC{{\mathcal{C}}}

\def\cE{{\mathcal{E}}}

\def\cG{{\mathcal{G}}}

\def\cK{{\mathcal{K}}}
\def\cL{{\mathcal{L}}}

\def\cN{{\mathcal{N}}}
\def\cO{{\mathcal{O}}}
\def\cP{{\mathcal{P}}}

\def\cU{{\mathcal{U}}}
\def\cV{{\mathcal{V}}}

\def\cX{{\mathcal{X}}}

\def\sfF{\mathsf{F}}

\def\E{{\mathbb{E}}}

\def\C{{\mathbb{C}}}
\def\R{\mathbb{R}}
\def\ind{{\mathbbm{1}}}

\newtheorem{remark}{Remark}

\newenvironment{proof outline}{\paragraph*{Proof Outline}}{\hfill$\IEEEQEDopen$}
\newcommand{\algorithmicinitialize}{\textbf{Initialize:}}
\newcommand{\INITIALIZE}{\item[\algorithmicinitialize]}

\def\act{\textup{act}}

\def\Npath{N_\textup{path}}

\def\Nsin{N_\textup{sin}}
\def\Nsub{N_\textup{sub}}
\def\Ptime{P_\textup{time}}
\def\Pfreq{P_\textup{freq}}

\newacronym{bs}{BS}{base station}
\newacronym{nr}{NR}{new radio}
\newacronym{gfra}{GFRA}{grant-free random access}
\newacronym{mimo}{MIMO}{multiple-input multiple-output}
\newacronym{siso}{SISO}{single-input single-output}
\newacronym{ofdm}{OFDM}{orthogonal frequency division multiplexing}
\newacronym{cir}{CIR}{channel impulse response}
\newacronym{dft}{DFT}{discrete Fourier transform}
\newacronym{fft}{FFT}{fast Fourier transform}
\newacronym{dct}{DCT}{discrete cosine transform}
\newacronym{pca}{PCA}{principal component analysis}
\newacronym{cs}{CS}{compressed sensing}
\newacronym{ml}{ML}{maximum-likelihood}
\newacronym{iid}{i.i.d.}{independent and identically distributed}
\newacronym{psd}{p.s.d.}{positive semidefinite}
\newacronym{amp}{AMP}{approximate message passing}
\newacronym{admm}{ADMM}{alternating direction method of multipliers}
\newacronym{omp}{OMP}{orthogonal matching pursuit}
\newacronym{bwl}{BWL}{block-wise linear}
\newacronym{bwp}{BWP}{block-wise polynomial}
\newacronym{wssus}{WSSUS}{wide-sense stationary and uncorrelated scattering}
\newacronym{wlrma}{WLRMA}{weighted low-rank matrix approximation}
\newacronym{qcqp}{QCQP}{quadratically constrained quadratic program}
\newacronym{sdp}{SDP}{semidefinite program}
\newacronym{lsfc}{LSFC}{large-scale fading coefficient}
\newacronym{mmv}{MMV}{multiple measurement vector}
\newacronym{smv}{SMV}{single measurement vector}
\newacronym{urllc}{URLLC}{ultra-reliable low-latency communication}
\newacronym{lasso}{LASSO}{least absolute shrinkage and selection operator}
\newacronym{mmse}{MMSE}{minimum mean-square error}
\newacronym{fr}{FR}{frequency range}
\newacronym{nnls}{NNLS}{non-negative least squares}
\newacronym{rrc}{RRC}{root-raised-cosine}
\newacronym{snr}{SNR}{signal-to-noise ratio}
\newacronym{roc}{ROC}{receiver operating characteristic}
\newacronym{evd}{EVD}{eigenvalue decomposition}
\newacronym{wrt}{w.r.t.}{with respect to}
\newacronym{rms}{RMS}{root-mean-square}
\newacronym{tdl}{TDL}{tapped delay line}
\newacronym{htx}{HTx}{hilly terrain}
\newacronym{tux}{TUx}{typical urban}
\newacronym{rax}{RAx}{rural area}
\newacronym{iot}{IoT}{internet-of-things}
\newacronym{em}{EM}{expectation maximization}
\newacronym{sa}{SA}{sequential approximation}

\makeglossaries

\begin{document}

\title{A Unified Activity Detection Framework for Massive Access:\! Beyond the Block-Fading Paradigm}

\author{Jianan~Bai and Erik~G.~Larsson
\thanks{This paper was presented in part at the  Asilomar SSC  2023 conference~\cite{conf}.}
\thanks{
The authors are with the Department of Electrical Engineering (ISY), Link\"oping University, 58183 Link\"oping, Sweden (email: jianan.bai@liu.se, erik.g.larsson@liu.se). 
This work was supported in part by Excellence Center at Link\"oping-Lund in Information Technology (ELLIIT), and by the Knut and Alice Wallenberg (KAW) foundation. 
The computations were enabled by resources provided by the National Academic Infrastructure for Supercomputing in Sweden (NAISS) partially funded by the Swedish Research Council through grant agreement no. 2022-06725.
}}

\maketitle

\begin{abstract}
	The wireless channel changes continuously with time and frequency and the block-fading assumption, which is popular in many theoretical analyses, never holds true in practical scenarios. 
	This discrepancy is critical for user activity detection in grant-free random access, where joint processing across multiple coherence blocks is undesirable, especially when the environment becomes more dynamic.
	In this paper, we develop a framework for low-dimensional approximation of the channel to capture its variations over time and frequency, and use this framework to implement robust activity detection algorithms. 
	Furthermore, we investigate how to efficiently estimate the principal subspace that defines the low-dimensional approximation. 
	We also examine pilot hopping as a way of exploiting time and frequency diversity in scenarios with limited channel coherence, and extend our algorithms to this case.  
	Through numerical examples, we demonstrate a substantial performance improvement achieved by our proposed framework.
\end{abstract}

\begin{IEEEkeywords}
	Massive access, activity detection, continuously varying channel, low-dimensional approximation, pilot hopping
\end{IEEEkeywords}

\section{Introduction}

Massive access features a scenario where a wireless network serves a very large number of users simultaneously \cite{chen2020massive}. 
Due to the limited radio resources, a considerable number of users need to communicate over the same time-frequency block of a size constrained by the channel \emph{coherence} length (i.e., the number of samples over which the channel remains relatively constant to allow for coherent signal processing).
Many applications in massive access, e.g., intelligent transportation systems and tactile internet, also require low access latency. 
Balancing these intertwined needs -- massive connectivity, efficient resource utilization, and minimal latency, is challenging.

In this paper, we focus on the initial stage in massive access, during which the active users inform the \gls{bs} of their needs for communication by sending a pilot sequence. 
Due to the limited channel coherence, users either need to reuse a set of mutually orthogonal pilots with contention or use unique, yet non-orthogonal pilots. 
We restrict our attention to the use of non-orthogonal pilots since they have shown superior detection performance compared to orthogonal pilots and are particularly suitable for short-packet transmission, which is generally the case in \gls{gfra} \cite{liu2018sparse}.
This process, commonly known as user \emph{activity detection}, has received much attention in recent years.

State-of-the-art activity detection schemes exploit the traffic sporadicity \cite{liu2018massive,fengler2021non}. 
That is, although the number of potential users can be exceedingly large, the number of simultaneously active users is comparable to the channel coherence length. 
The majority of activity detection methods fall into two primary categories: \gls{cs}-based and covariance-based approaches. 
Consider a network with a multi-antenna \gls{bs} and single-antenna users, the \gls{cs}-based approaches are motivated by the linear measurement model $\bY=\bPhi\bX + \bW$ that appears in pilot transmission, where each column of $\bY$ is the received signals at one \gls{bs} antenna, each column of $\bPhi$ represents the pilot sequence of a user, each row of $\bX$ represents the effective channel gains (assumed to be \emph{constant}) from a user to all \gls{bs} antennas, and $\bW$ represents additive noise. 
Due to the sporadic traffic, $\bX$ is row-sparse and can be accurately recovered under certain conditions, and activity detection can be performed by thresholding the norm of each row in the recovered $\bX$.
The \gls{cs}-based algorithms range from conventional optimization-based methods \cite{djelouat2022spatial,leinonen2019compressed} to statistical methods using, for example, \gls{amp} \cite{liu2018massive}.
There are several advantages of the \gls{cs}-based approaches. 
First, they directly provide the channel estimates as a byproduct. 
Second, additional side information (e.g., temporal correlation \cite{wang2022exploiting}) can be incorporated, especially for the statistical methods. 
Third, although the increase in the number of \gls{bs} antennas can improve detection performance in \gls{mimo} systems, the \gls{cs}-based approaches can work for a moderately small number of antennas, enabling flexible deployments.
However, the detection capabilities of \gls{cs}-based approaches are fundamentally limited by the need to recover all non-zero elements in $\bX$ -- we usually cannot accurately detect more active users than the pilot length without additional side information. 
Furthermore, as observed in \cite{fengler2021non}, the \gls{amp} algorithm may exhibit random, non-convergent behaviors in certain application regimes.

The covariance-based approach \cite{fengler2021non}, which originates from the support recovery problem in sparse Bayesian learning \cite{jin2013support}, takes another perspective when solving the activity detection problem. 
Specifically, when the non-zero elements in $\bX$ are statistically independent (let them also be zero-mean and unit-variance for simplicity), as the number of \gls{bs} antennas $M$ increases, the sample covariance matrix $\bY\bY^\herm/M$ approaches the true covariance matrix $\bPhi\bD_{\ba}\bPhi^\herm + \sigma^2\bI $, which is parameterized by the binary vector of user activities  $\ba$, with noise variance $\sigma^2$. 
An efficient coordinate descent algorithm is developed in \cite{fengler2021non} to recover $\ba$ by matching these two covariance matrices \gls{wrt} the log-determinant divergence -- a direct consequence of a \gls{ml} estimation of $\ba$.
Different from the \gls{cs} approaches, the covariance-based one does not directly provide channel estimates (which may be acquired after activity detection, using, for example, linear \gls{mmse} estimation), and it becomes less straightforward to incorporate statistical side information (although not impossible, as showcased in \cite{jiang2021ml}). 
However, the covariance-based approach can identify a considerably larger number of active users than the \gls{cs}-based ones in large, albeit finite, antenna regimes; the fundamental scaling law is formally substantiated in \cite{fengler2021non}. 

Despite the promising detection performance showcased in various scenarios, current activity detection schemes (e.g., \cite{liu2018massive,fengler2021non,djelouat2022spatial,wang2022exploiting,jiang2021ml}) are predominantly restricted to an ideal block-fading model, where the channel is assumed to be constant during pilot transmission to make the linear measurement model $\bY=\bPhi\bX + \bW$ valid. 
This block-fading assumption can be formally justified by the concept of coherence block (a typical definition is a time-frequency block with up to $\pi$ phase shifts \cite[Ch. 2.1]{marzetta2016fundamentals}) and the sampling theorem, only when joint processing or coding across a sufficiently large number of blocks is feasible.
Nevertheless, interpolation may not be possible in the context of activity detection, wherein instantaneous processing is anticipated within a block or a few blocks.
The situation is most challenging in highly dynamic IoT environments associated with emerging 5G scenarios, where higher operating frequencies, larger sub-carrier spacings, and potentially high mobility need to be considered.

Some efforts have been made to address the channel variations in activity detection in \gls{ofdm} systems. 
In \cite{jiang2022massive}, sub-carriers are partitioned into sub-blocks, and the channel gain is assumed to change linearly within each sub-block; consequently, an \gls{amp}-based detection algorithm was developed. 
The method in \cite{jiang2022massive} was further refined in \cite{scharf2023user} by approximating the channel variations in each sub-block as a low-order polynomial. 
The covariance-based approach was extended for \gls{ofdm} systems in \cite{jiang2022statistical} by assuming \gls{iid} channel taps in the delay domain and capitalizing on the \gls{dft} structure present in the \gls{ofdm} symbols.
A similar idea was explored in \cite{zhu2022ofdm} by using the \gls{dct} representations.
All these approximation models, including the simplest block-fading model, as we will show later, can be unified from a dimensionality reduction perspective.
Another direction to combat the channel variations in activity detection is to use pilot hopping to explore extra time and/or frequency diversities, as demonstrated in \cite{de2017random,becirovic2019detection}.

\subsection{Contributions and Organization of the Paper}

\noindent\textbf{1) Robust Activity Detection:}

We consider a general channel model for activity detection in Section \ref{sec: system model}, which allows the channel coefficients to change symbol-by-symbol, departing from the commonly assumed block-fading model in the literature. 
We present a dimensionality-reduction framework for channel variations that generalizes various approaches in, for example, \cite{jiang2022massive,scharf2023user,jiang2022statistical,zhu2022ofdm}. 
By leveraging the low-dimensional structure, we implement a modified covariance-based activity detection algorithm that is robust under highly varying channels.

\noindent\textbf{2) Low-Dimensional Structure Learning:}

The proposed framework relies on the knowledge of a low-dimensional structure of the channel, which is generally not known a priori.
We investigate the learning of the low-dimensional structure in Section \ref{sec: learn channel covariance}, by allowing the users to transmit additional, \emph{dedicated} all-one pilots.
In this case, the low-dimensional structure can be accurately learned by solving a standard low-rank matrix approximation problem, without the need for knowing the user activities.

We also explore \emph{joint} estimation of the low-dimensional structure and detection of the user activities, by reusing the same received signals.
For this joint problem, we consider an alternating procedure, where we iteratively update the estimates of user activities and the low-dimensional structure.
Particularly, the structure learning is formulated as a \gls{wlrma} problem, where the weight matrix is determined by the (estimated) user activities.
However, we observe that this approach performs poorly due to the inherent ill-conditioning of the joint problem.
Details on this approach are provided in the Appendix.

\noindent\textbf{3) Pilot Hopping:}

We examine the pilot hopping scheme and combine it with the activity detection algorithm in Section \ref{sec: pilot hopping}. 
We develop a new algorithm for hopping pattern generation, inspired by the configuration model for random graphs \cite[Ch. 5.3]{latora2017complex}. 
The proposed hopping pattern generation algorithm significantly outperforms the random hopping pattern generation methods in \cite{de2017random,becirovic2019detection}. 
Furthermore, we use pilot hopping to reduce the overhead associated with learning the low-dimensional structure of the channels using dedicated pilots.

\textbf{Remark:}
We presented the dimensionality-reduction framework for activity detection in the conference paper \cite{conf}. The learning of the low-dimensional structure and the pilot hopping scheme are new contributions introduced in this paper.

\subsection{Notation}

Vectors are denoted by boldface lowercase letters, $\bx$, matrices by boldface uppercase letters, $\bX$, with determinant $|\bX|$, and sets by calligraphic letters, $\cX$, with cardinality $|\cX|$. 
$[N]$ denotes the set $\{1,\cdots,N\}$.
$(\cdot)^\transp$, $(\cdot)^\herm$, $(\cdot)^*$, and $(\cdot)^{-1}$ denote transpose, conjugate transpose, complex conjugate, and inverse, respectively. 
$\bI_N$, $\bone_N$, and $\bzero_N$ denote respectively the identity matrix, the all-one vector, and the all-zero vector of size $N$ (omitted when the size is obvious). $\E[\cdot]$ denotes the statistical expectation. $\ind\{\cdot\}$ is the indicator function, which equals to $1$ for true propositions and $0$ otherwise. The multivariate circularly symmetric complex Gaussian distribution with covariance $\bR$ is denoted by $\cn{\bzero}{\bR}$. $\bD_{\bx}$ denotes a diagonal matrix with $\bx$ on its diagonal. 
%$[\bx]_i$ and $[\bX]_{i,j}$ represent the $i$th entry in $\bx$ and the $(i,j)$th entry in $\bX$, respectively. 
$\R_+$, $\C$, and $\mathbb{S}_+$ denote the spaces of non-negative numbers, complex numbers, and \gls{psd} matrices. 
$\|\cdot\|$ denotes the norms.
%$\|\cdot\|_\sfF$ is the Frobenius norm.

\section{System Model}

\label{sec: system model}

We consider a single-cell system with an $M$-antenna \gls{bs} and $K$ single-antenna users. 
Each user, $k\in[K]$, is pre-assigned a unique pilot sequence of length $L$, denoted by $\bphi_k \defeq [\phi_{k1},\cdots,\phi_{kL}]^\transp$, which is normalized to have unit average energy per symbol, i.e., $\|\bphi_k\|_2^2=L$. 
Our focus lies primarily in the regime where $L\ll K$, and therefore, pilot sequences allocated to distinct users are mutually non-orthogonal in general, i.e., $\bphi_i^\herm\bphi_j\neq 0$ when $i\neq j$.

In a given communication round, the active users, denoted by $\cK_\act$, with $|\cK_\act| \ll K$ due to the sporadic traffic, transmit their pre-assigned pilots synchronously on a time-frequency block of size $T \times F$ (i.e., a resource block with $T$ \gls{ofdm} symbols and $F$ sub-carriers). 
We assume $L=TF$ for simplicity. At the $m$th antenna, denoting by $h_{lkm}$ the small-scale fading coefficient experienced by the $l$th pilot symbol sent from user $k$, the received pilot signal is given by
\begin{equation}
\label{eq: received signal}
	\by_m = \sum_{k\in[K]} a_k \sqrt{\beta_k} \bD_{\bh_{km}}\bphi_k + \bw_m,
\end{equation}
where $a_k\in\{0,1\}$ represents the activity of user $k$, $\beta_k$ is the received signal strength (i.e., the product of the \gls{lsfc} and the transmit power), $\bh_{km} \defeq [h_{1km},\cdots,h_{Lkm}]^\transp$ is the channel vector from user $k$ to antenna $m$, and $\vect{w}_m\sim\cn{\vect{0}}{\sigma^2\mat{I}}$ is additive noise that is independent across antennas.

Different from the block-fading model where the channel coefficients are assumed to be constant during pilot transmission, i.e., $\bh_{km}=\overline{h}_{km}\bone_L$ for some scalar-valued $\overline{h}_{km}$, the received signal model \eqref{eq: received signal} allows the transmitted symbols to experience different, static channels. 
Such a model can be obtained by prepending sufficiently long cyclic prefixes to the \gls{ofdm} symbols, because the cyclic convolution operation is diagonalized in the coordinate system defined by columns of the \gls{dft} matrix, as elucidated in, for example, \cite[Ch. 7.1]{bjornson2024introduction}.

\begin{remark}
	\label{remark: static channel}
	Our model assumes that the channel is static within a symbol, such that the effect of the channel for each symbol is to multiply the symbol with a complex number; in particular, there is no inter-symbol or inter-carrier interference. 
	In practice, the channel is time-varying, and this model is, strictly speaking, an approximation.
	However, the channel generally varies slowly compared to the \gls{ofdm} symbol duration, and this approximation is usually sufficiently good and used in real-world algorithm implementations of data transmission with \gls{ofdm}.

\end{remark}

\section{Activity Detection with Dimensionality-Reduced Channel Variations}

\label{sec: dimensionality reduction}

The underlying physical propagation model generally permits a low-dimensional structure in the channel vectors that we can exploit. 
Mathematically, a low-dimensional approximation of $\bh_{km}$ can be expressed as
\begin{equation}
\label{eq: channel-approx}
	\bh_{km} \approx \bG \btheta_{km} = \sum_{n\in[N]} \theta_{nkm}\bg_n,
\end{equation}
where $\vect{\theta}_{km}=[\theta_{1km},\cdots,\theta_{Nkm}]^\transp$ is a random vector with \gls{iid} entries, $\bG \defeq [\bg_1,\cdots,\bg_N]\in\C^{L\times N}$ is a deterministic basis matrix (the basis vectors $\{\bg_n\}$ may have different lengths), and we refer to $N$ as the approximation \emph{order}. 
The intuition behind \eqref{eq: channel-approx} is that all the channel vectors $\{\vect{h}_{km}\}$ approximately lie in an $N$-dimensional subspace, with $N\ll L$, spanned by $\{\vect{g}_n\}$, and the stochastic nature of $\vect{h}_{km}$ is encapsulated within a significantly reduced representation $\vect{\theta}_{km}$.

The assumed low-dimensional structure is motivated by the fact that even if the channel varies substantially within a block, the fading experienced by adjacent pilot symbols can still exhibit a strong statistical correlation. 
This can be seen as a generalization of the coherence block, nominally defined in a deterministic sense (one typical definition is a time-frequency block with up to $\pi$ phase shifts), to a statistical viewpoint.

\begin{remark}
	\label{remark: low-dimensional model}
	The concept of low-dimensional approximation by exploring  statistical correlation as such is not new; it originates from the idea of \gls{pca} \cite{pearson1901pca}.
	The existing works \cite{jiang2022massive,scharf2023user,jiang2022statistical,zhu2022ofdm} can be seen as special cases of the model in \eqref{eq: channel-approx}. 
	Previously, the channel subspace was mainly exploited in the spatial domain (e.g., \cite{haghighatshoar2016massive}), instead of \emph{over time and frequency} as in our work.
\end{remark}

\subsection{Existing Low-Dimensional Models}
\label{sec: existing low-dimensional models}

The block-fading model, for which $\bG=\bone_{L}$, is the simplest example of \eqref{eq: channel-approx}. 
The block-fading assumption can be justified by the sampling theorem, which allows us to interpolate back to the continuous channel variations by taking one sample from each coherence block when the number of samples is sufficiently large. 
However, typical applications of \gls{gfra} require low latency, which limits the number of blocks that can be jointly processed.
The problem becomes more challenging when substantial channel variations are present due to a large excess delay and high mobility in dynamic IoT environments.

Recently, research efforts have been made for activity detection in wideband systems, especially in \gls{ofdm} systems. We can classify these approaches into two primary categories, and both can be seen as special cases of \eqref{eq: channel-approx}:

\noindent 1)
The first category of approaches approximates the channel variations in a block-wise manner. 
Specifically, the OFDM sub-carriers are divided into multiple sub-blocks. 
The variations within each sub-block are approximated as a linear function in \cite{jiang2022massive}, or as a low-degree polynomial in \cite{scharf2023user}. 
These models correspond to a block-diagonal basis matrix $\bG$. 
As an example, the \gls{bwl} model has $\bG = \bdiag(\bG_0,\cdots,\bG_0)$ with $N/2$ diagonal blocks (assume $N/2$ is an integer). 
Each diagonal block $\bG_0$ is a $(L/N)\times 2$ matrix, with the first column being an all-one vector accounting for the mean value, and the entries in the second column are equally spaced and centered at zero, representing the linear variations. 
The \gls{bwp} model can be constructed similarly by incorporating additional columns in the diagonal blocks accounting for the higher degrees in the polynomial.

\noindent 2)
Another direction of work utilizes the transform-domain representations. 
Particularly, in \gls{ofdm} systems, the channel frequency responses are Fourier transforms of \glspl{cir}.
Since the \glspl{cir} are generally dominated by a few strong channel taps, the basis matrix can be selected as the corresponding columns in the \gls{dft} matrix~\cite{jiang2022statistical} when assuming that these dominant channel taps are independently distributed. 
A similar method in \cite{zhu2022ofdm} is to use the columns in the \gls{dct} matrix instead.

\subsection{A Viewpoint from Low-Rank Covariance Approximation}
\label{subsec: low-rank cov}

To further motivate the dimensionality reduction techniques and to unify those low-dimensional channel models, we provide another viewpoint here. 
We assume that the channel between each user-antenna pair is modeled by the correlated Rayleigh fading that is stationary across different user-antenna pairs.
In this case, $\bh_{km}\sim \cC\cN(\bzero,\bR)$ with the same $L\times L$ covariance matrix\footnote{We have two different types of covariance matrix in this paper: the \emph{channel} covariance matrix $\bR \defeq \E[\bh\bh^\herm]$ and the \emph{signal} covariance $\bSigma \defeq \E[\by\by^\herm]$. They are not to be confused with each other or with the \emph{spatial} covariance matrix.} $\bR$ for all $k\in[K]$ and $m\in[M]$. 
The channel covariance $\bR$ is assumed to have rank $N$, or to have a good rank-$N$ approximation.
Then, we can obtain a Gramian factorization $\bR = \bG\bG^\herm$, where the $L\times N$ matrix $\bG$ serves as the basis matrix in \eqref{eq: channel-approx}. 
Additionally, under the Rayleigh fading assumption, the random vector $\btheta_{km}$ has the distribution $\cC\cN(\bzero,\bI_N)$.
The aforementioned low-dimensional channel approximation approaches, by imposing the particular structures in $\bG$, equivalently assume corresponding special structures in the channel covariance matrix $\bR$. 

\begin{remark}
\label{remark: spatial stationarity}
For simplicity, we assume that the small-scale fading is spatially stationary to reduce the number of unknowns in the channel covariance estimation. This assumption can be justified for example if the users are in the far-field of the array, and move around the area over a sufficiently long time window. However, in practice, users located in different areas may certainly experience non-stationarities. 
To model this, one could for example group the users based on their locations, with location-dependent channel covariances.
We have to leave this as a topic for possible future work.
\end{remark}

\subsection{Covariance-Based Activity Detection}
\label{sec: covariance approach}

We assume that the channel covariance $\bR$ is known, has rank $N$, and can be factorized into the Gramian form $\bR=\bG\bG^\herm$, as discussed in Section \ref{subsec: low-rank cov}.
(If $\bR$ is not perfectly known, we replace it with an estimate. A discussion on estimating a low-rank $\bR$ will be provided in Section~\ref{sec: learn channel covariance}.) 
We aim to detect the user activities $\{a_k\}$ by exploiting this low-rank structure. 
Although there are various algorithms for activity detection based on \gls{cs}, we will focus on the covariance approach proposed in \cite{fengler2021non}. 
This approach does not directly produce channel estimates but can potentially detect many more active users than the \gls{cs}-based approaches when using a large antenna array in massive \gls{mimo} systems.

The covariance approach for activity detection is developed from a (relaxed) \gls{ml} estimation of $\bgamma \defeq [\gamma_1,\cdots,\gamma_K]^\transp$, with $\gamma_k\defeq a_k\beta_k$. 
By assuming \gls{iid} $\{\bh_{km}\}$ with distribution $\cC\cN(\bzero,\bR)$, the received signals $\{\by_m\}$ become \gls{iid} with distribution $\cC\cN(\bzero,\bSigma_{\bgamma})$ and the covariance matrix 
\begin{equation}
\label{eq: cov}
	\bSigma_{\bgamma} \defeq \E[\by_m\by_m^\herm] = \sum_{k\in[K]}\gamma_k\bS_k\bS_k^\herm + \sigma^2\bI,
\end{equation}
where $\bS_k\defeq \bD_{\bphi_k}\bG$ is the effective pilot (matrix) of user $k$. In the block-fading case where $\bG = \bone$, we have $\bS_k=\bphi_k$, and each active user contributes a rank-one component in the signal covariance. However, in general cases with varying channels, the contribution from each active user becomes a rank-$N$ component that complicates the problem. 

After some rescaling and removal of constant terms, the negative log-likelihood function of $\bgamma$ given $\{\by_m\}$ is 
\begin{equation}
\label{eq: ML objective}
	f(\bgamma) \defeq \log|\bSigma_{\bgamma}| + \tr(\bSigma_{\bgamma}^{-1}\widehat{\bSigma}),
\end{equation}
\begin{flalign}
\label{eq: sample covariance}
	\mbox{where} && \widehat{\bSigma} \defeq \frac{1}{M}\sum_{m\in[M]}\vect{y}_m\vect{y}_m^\herm &&
\end{flalign}
is the sample covariance matrix of the received signals.
Notice that $\widehat{\bSigma}$ is a sufficient statistic for the estimation of $\bgamma$, and the \gls{ml} formulation can also be interpreted as a covariance matching problem by minimizing the log-determinant divergence between $\widehat{\bSigma}$ and $\bSigma_{\bgamma}$.
The binary constraint $a_k\in\{0,1\},\forall k$, renders the original \gls{ml} problem combinatorial, causing the complexity to grow exponentially with $K$. 
The search space can be relaxed to $\bgamma\in\Gamma$, where $\Gamma$ represents the box constraint $\{\bgamma\in\R^K: 0\leq \gamma_k\leq \beta_k,\forall k\}$ when the \glspl{lsfc} $\{\beta_k\}$ are known, and represents the non-negative orthant $\R_+^K$ otherwise. 
The relaxed \gls{ml} estimation of $\bgamma$ is then given by 
\begin{equation}
	\widehat{\bgamma}_\textup{ML} = \argmin_{\bgamma \in \Gamma} f(\bgamma).
	\tag{P0}
	\label{P0}
\end{equation}

An efficient coordinate descent algorithm is developed in \cite{fengler2021non} for block-fading channels. In each (inner) iteration of the coordinate descent algorithm, we pick a coordinate (user) $k$ based on some pre-determined schedule and make the update
\begin{equation}
	\bgamma \leftarrow \bgamma + d^*\bee_k,
\end{equation}
where $\bee_k$ is the $k$th column of $\bI_K$, and
\begin{equation}
\label{eq: cost 1d}
	d^* = \argmin_{d \in [-\gamma_k, \infty)} f(\bgamma + d\bee_k).
\end{equation}
When $N = 1$, changing $d$ results in a rank-one update in the covariance matrix, and the optimal $d^*$ can be obtained in closed form~\cite{fengler2021non}. 
An extension of this approach was proposed in \cite{jiang2022statistical}, and we can apply it to the general case when $N > 1$.
For completeness, we briefly present the development of the approach in \cite{jiang2022statistical} in the following.

By applying Sylvester's determinant identity we obtain
\begin{equation*}
	\log |\bSigma_{\bgamma} + d\mat{S}_k\mat{S}_k^\herm | = \log|\bSigma_{\bgamma}| + \log|\mat{I} + d\bA_k|,
\end{equation*}
where $\bA_k=\bS_k^\herm\bSigma_{\bgamma}^{-1}\bS_k$, and Woodbury's matrix identity gives
\begin{equation*}
\begin{aligned}
	(\bSigma_{\bgamma} + d\mat{S}_k\mat{S}_k^\herm)^{-1} = \bSigma_{\bgamma}^{-1} - d\bSigma_{\bgamma}^{-1}\mat{S}_k(\mat{I} + d\bA_k)^{-1}\mat{S}_k^\herm\bSigma_{\bgamma}^{-1}.
\end{aligned}
\end{equation*}
After the \gls{evd} $\bA_k=\bV_k\bD_{\blambda_k}\!\!\bV_k^\herm$  with $\blambda_k = [\lambda_{k1},\cdots,\lambda_{kN}]^\transp$, the cost function in \eqref{eq: cost 1d} becomes
\begin{equation}
\begin{aligned}
\label{eq:cost-1d}
	\cL_k(d) \defeq& f(\bgamma\!+\!d\bee_k) - f(\bgamma)\\
	=& \log|\mat{I}\! +\! d\mat{D}_{\vect{\lambda}_k}|\! -\! d \tr \left((\mat{I}\! +\! d\mat{D}_{\vect{\lambda}_k})^{-\!1}\mat{\Xi}_k \right)\\
	=& \sum_{n\in[N]}\!\!\left(\log(1\!+\!d\lambda_{kn}) \!-\! \frac{d\xi_{kn}}{1\!+\!d\lambda_{kn}} \right)
\end{aligned}
\end{equation}
with $\mat{\Xi}_k = \mat{V}_k^\herm\mat{S}_k^\herm\bSigma_{\vect{\gamma}}^{-1}\widehat{\bSigma}\bSigma_{\vect{\gamma}}^{-1}\mat{S}_k\mat{V}_k$ and $\xi_{kn}=[\mat{\Xi}_k]_{n,n}$.

The optimal $d^*$ can be obtained by comparing the cost value for all feasible stationary points of \eqref{eq:cost-1d} and boundary points.
To find the stationary points of \eqref{eq:cost-1d}, where the derivative $\cL_k'(d)$ equals zero, one needs to solve 
\begin{equation}
\label{eq:derivative}
	\cL_k'(d) = \sum_{n\in[N]} \left( \frac{\lambda_{kn}}{1 + d\lambda_{kn}} + \frac{\xi_{kn}}{(1+d\lambda_{kn})^2} \right) = 0.
\end{equation}
Multiplying  both sides of the second equality in \eqref{eq:derivative} by $\prod_{n\in[N]}(1+d\lambda_{kn})^2$,  the stationary points can also be represented as the solutions to
\begin{equation}
\label{eq:polynomial}
	\sum_{n=1}^N \left(\lambda_{kn}(1+d\lambda_{kn}) +\xi_{kn}\right) \prod_{j\neq n}(1+d\lambda_{kj})^2 = 0,
\end{equation}
which is a polynomial equation with order $2N-1$. No explicit formula exists for $N\geq 3$, while efficient algorithms for the real-root isolation of high-order polynomials were developed \cite{kobel2016computing,mcnamee2013numerical}. 
Alternatively, as suggested in \cite{chen2021sparse}, a one-dimensional search can be employed to minimize \eqref{eq:cost-1d} directly.

The procedure above is repeated for $I$ (outer) iterations.
We summarize the overall procedure in Algorithm \ref{alg: activity detection}. 

\textbf{Complexity:}
The runtime complexity of Algorithm \ref{alg: activity detection} is dominated by: 
1) the matrix multiplications in steps \ref{CD step: EVD}, \ref{CD step: V-tilde}, \ref{CD step: xi}, and \ref{CD step: update covariance}, which require $\cO(L^2 N)$ arithmetic operations; 
2) the \gls{evd} in step \ref{CD step: EVD}, which requires $\cO(N^3)$ arithmetic operations; 
and 3) finding the minimizer of $\cL_k(d)$ in step \ref{CD step: 1d minimization}. 
Finding the roots in \eqref{eq:derivative} has complexity $\cO(N^3)$.
Alternatively, we can use golden section search and parabolic interpolation to perform a one-dimensional search whose complexity does not scale with $N$ \cite{brent2013algorithms}.
Overall the algorithm has a computational complexity of order $\cO(IKN(L^2+N^2))$.

\begin{algorithm}[t]
	\caption{Activity Detection}
	\begin{algorithmic}[1]
		\label{alg: activity detection}
		\REQUIRE sample covariance $\widehat{\bSigma}$, and effective pilots $\{\bS_k\}$
		\INITIALIZE $\gamma_k \gets 0, \forall k$ and $\bSigma_{\bgamma}^{-1} \gets \sigma^{-2}\bI$
		\FOR{$i=1,2,\cdots,I$}
		\STATE Generate $\cK$ by randomly permuting $[K]$ 
		\FOR{$k$ taken from $\cK$ by order}
		\STATE  $\bV\bD_{\blambda}\bV^\herm \xleftarrow{\textup{EVD}} \bS_k^\herm\bSigma_{\bgamma}^{-1}\bS_k $ \label{CD step: EVD}
		\STATE $\widetilde{\bV} \gets \bSigma_{\bgamma}^{-1}\bS_k\bV$ \label{CD step: V-tilde}
		\STATE $\forall n$: $\xi_n \gets [\widetilde{\bV}^\herm\widehat{\bSigma}\widetilde{\bV}]_{n,n}$ \label{CD step: xi}
		\STATE $d^* = \argmin \cL_k(d)$ for $d \in [-\gamma_k,\gamma_{\max}-\gamma_k]$\\
			($\gamma_{\max}$ can be selected as an estimate of $\max \{\beta_k\}$) \label{CD step: 1d minimization}
		\STATE $\bSigma_{\bgamma}^{-1} \gets \bSigma_{\bgamma}^{-1} - d^*\widetilde{\bV}(\mat{I} + d^*\bD_{\blambda})^{-1}\widetilde{\bV}^\herm$ \label{CD step: update covariance} 
		\STATE $\gamma_k \gets \gamma_k + d^*$
		\ENDFOR
		\ENDFOR
		\ENSURE activity estimate $\bgamma = [\gamma_1,\cdots,\gamma_K]^\transp$
	\end{algorithmic}
\end{algorithm}

\section{Learning the Low-Dimensional Structure}

\label{sec: learn channel covariance}

In Section \ref{sec: dimensionality reduction}, we assumed that the channel vectors lie in a low-dimensional linear subspace as the result of a low-rank channel covariance matrix. 
When the channel covariance matrix $\bR$ is known, the optimal approximation that minimizes the approximation order can be obtained by projecting the channel vectors onto their principal subspace. 
However, in practical wireless environments, the channel covariance will not be perfectly known and may continuously change. 

A naive approach is to first estimate the channels and use these channel estimates to form a channel covariance estimate. 
However, this approach fails when we allow symbol-to-symbol channel variations, as we cannot estimate $LKM$ unknowns in $\{\bh_{km}\}$ using the $LM$ observations from $\{\by_m\}$. 
While it may seem appealing to use the low-dimensional channel model in \eqref{eq: channel-approx} to facilitate channel estimation by reducing the number of unknowns to $NKM$, it is crucial to recognize that assuming a low-dimensional structure for channel estimation already confines the channel estimates within that pre-determined subspace, making it impossible to update this subspace.

In what follows, we consider learning the low-dimensional channel structure directly from the received signals.

Notice that the covariance matrix of the received signal $\by_m$ in \eqref{eq: cov} can be re-written as
\begin{equation}
\label{eq: masked covariance}
	\bSigma_{\bgamma,\bR} =  \bC(\bgamma)\odot \bR + \sigma^2\bI,
\end{equation}
\begin{flalign}
\label{eq: C matrix}
	\mbox{where} && \bC(\bgamma) \defeq \sum_{k\in[K]} \gamma_k\bphi_k\bphi_k^\herm, &&
\end{flalign}
and $\odot$ represents the Hadamard (element-wise) matrix product.
It might appear attempting to jointly estimate $\bgamma$ and $\bR$. 
For instance, the joint \gls{ml} estimation can be formulated as 
\begin{equation}
\label{eq: joint ML}
\begin{aligned}
	\min_{\bgamma\in\Gamma, \bR\in\mathbb{S}_+^L}~&\log|\bSigma_{\bgamma,\bR}| + \tr(\bSigma_{\bgamma,\bR}^{-1}\widehat{\bSigma})\\
	\operatorname{s.t.}~~~~& \rank(\bR) \leq N.
\end{aligned}
\end{equation}
Assuming an estimate of $\bR$ from previous transmissions, denoted by $\widehat{\bR}$, we may consider the following alternating procedure to estimate $\bgamma$ and update $\widehat{\bR}$: 
1) find an activity estimate $\widehat{\bgamma}$ while fixing $\widehat{\bR}$ by running the coordinate descent algorithm in Section \ref{sec: covariance approach}, then 
2) re-estimate the channel covariance while keeping $\widehat{\bgamma}$ fixed (to be discussed in the Appendix). These two steps can be repeated for several iterations.

However, the joint estimation problem appears to be ill-conditioned, and our experiments have not yielded much success.
We suspect this is because the estimation of $\bgamma$ and $\bR$ are two conflicting objectives that require pilots with contradicting properties. 
When estimating $\bgamma$, one generally prefers pilot sequences with low cross-correlation to better distinguish users by their unique pilots. 
However, sending pilots with low cross-correlation typically results in a diagonally dominant $\bC(\bgamma)$ in \eqref{eq: C matrix}, while the off-diagonal elements in $\bSigma_{\bgamma,\bR}$ have low magnitudes and become sensitive to noise and estimation error of $\bgamma$. 
On the other hand, when the focus shifts to the estimation of $\bR$, using the all-one sequence,  i.e., $\bphi_k=\bone$, becomes an obvious choice, as the weight matrix becomes $\bC(\bgamma) = (\sum_{k\in[K]}\gamma_k) \bone\bone^\transp$, reducing the Hadamard product with $\bC(\bgamma)$ to scalar multiplication with $\sum_{k\in[K]}\gamma_k$.
However, by forcing all users to send the same pilot sequence, user identification becomes impossible.

Exploring an optimal tradeoff between activity detection and low-dimensional structure learning through pilot design presents a formidable challenge. 
We will not investigate further in this direction and will instead restrict our focus to the use of dedicated all-one pilot sequences for channel covariance estimation. 
When using the all-one pilots, the learning of low-dimensional channel structure is straightforward.
We first perform the \gls{evd} on $\widehat{\bSigma}-\sigma^2\bI$ and take the $N$ dominant eigenvalues $\{\varsigma_n\}$ and the corresponding eigenvectors $\{\bmu_n\}$.
The $n$th basis vector of the low-dimensional space is taken as
\begin{equation}
\label{eq: basis-estimate}
	\bg_n = \sqrt{\frac{L \varsigma_n}{\sum_{n'\in[N]} \varsigma_{n'}}} \bmu_n,
\end{equation}
where the scaling factor makes the small-scale fading coefficients unit-variance, i.e., $h_{nkm}\sim\cn{0}{1}$.
The corresponding channel covariance is $\bR = \sum_{n\in[N]} \bg_n \bg_n^\herm$.
Notice that we do not need to know the user activities when estimating $\{\bg_n\}$ since the user activities introduce only a scaling factor which will be removed by the rescaling in \eqref{eq: basis-estimate}.

The drawback of using dedicated all-one pilots for low-dimensional structure learning is that the users have to make additional transmissions, as those pilots cannot be reused for activity detection. 
However, as will be elaborated in Section~\ref{sec: complete framework}, this overhead can be substantially reduced when combined with pilot hopping.

\section{Extension: Pilot Hopping}

\label{sec: pilot hopping}

\begin{figure}
	\centering
	\includegraphics[width=0.8\linewidth]{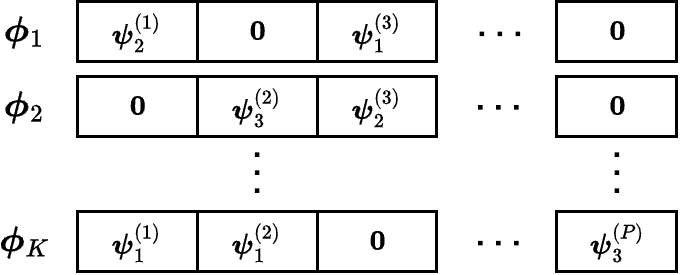}
	\caption{Pilot hopping as integrating a special structure in pilots.}
	\label{fig: pilot-hopping}
\end{figure}

Another scheme to combat the limited channel coherence in activity detection is to apply pilot hopping, which has been considered in \cite{de2017random}. 
Specifically, the $L$ samples in a channel block are partitioned into $P$ sub-blocks of size $\tau$, with $L=P\tau$ for simplicity. 
 Within each sub-block $p\in [P]$, $J$ sequences $\bpsi_1^{(p)},\cdots,.\bpsi_J^{(p)}$ of length $\tau$ are generated, and we refer to $\bpsi_j^{(p)}$ as the $j$th sub-pilot in the $p$th sub-block. 
 Each user $k$ is assigned a unique hopping pattern $\bz_k \defeq [z_k^{(1)},\cdots,z_k^{(P)}]^\transp$, where $z_k^{(p)} \in \{0\}\cup[J]$.
 If $z_k^{(p)}\in[J]$, user $k$ transmits the $z_k^{(p)}$th sub-pilot in the $p$th sub-block when it is active; no pilot transmission if $z_k^{(p)}=0$. 
 By defining the $J\times K$ sequence selection matrices $\{\bU^{(p)}\}$ with $[\bU^{(p)}]_{j,k} = \ind\{z_k^{(p)}=j\}$, the pilot matrix can be written as 
\begin{equation}
\label{eq: pilot with hopping}
	\bPhi = \begin{bmatrix}
		\bPhi^{(1)}\\
		\vdots\\
		\bPhi^{(P)}
	\end{bmatrix},\quad 
	\bPhi^{(p)} \defeq \bPsi^{(p)}\bU^{(p)},
\end{equation}
where $\bPsi^{(p)} \defeq [\bpsi_1^{(p)},\cdots,\bpsi_J^{(p)}]$. 
From \eqref{eq: pilot with hopping}, we can observe that the pilot hopping scheme can be seen as introducing a special structure in the pilot matrix; see Fig. \ref{fig: pilot-hopping}.

In \cite{de2017random}, the channel is assumed to be quasi-static so that it stays constant within each sub-block and changes independently across sub-blocks. 
The sub-pilots are made mutually orthogonal (i.e., $J=\tau$) and the users are distinguished by their unique hopping patterns. 
A practical activity detection algorithm for this scheme is proposed in \cite{becirovic2019detection}, relying on the channel hardening and the favorable propagation properties of massive \gls{mimo}. 
More specifically, as shown in \cite{becirovic2019detection}, when the number of receive antennas $M\rightarrow \infty$, the received energies on $\bpsi_j^{(p)}$ has the form 
\begin{equation}
	e_j^{(p)} \rightarrow \underbrace{[\ind\{z_1^{(p)}=j\}\tau\beta_1,\cdots,\ind\{z_K^{(p)}=j\}\tau\beta_K]}_{\defeq (\vect{\omega}_j^{(p)})^\transp }\ba.
\end{equation}

By stacking $\{e_j^{(p)}\}$ into a vector $\vect{e}$, the relation can be written as $\vect{e} \xrightarrow{M\rightarrow \infty} \mat{\Omega}\ba$, where $\mat{\Omega}$ has rows $\{(\vect{\omega}_j^{(p)})^\transp\}$. 
The $P\tau\times K$ matrix $\mat{\Omega}$ can be seen as a sparse measurement (sensing) matrix, and a corresponding \gls{cs} problem can be formulated to recover the user activities $\ba$:
\begin{equation}
\label{eq: lasso}
	\min_{\ba \in[0,1]^K}~\|\mat{\Omega}\ba - \vect{e}\|_2^2 + \lambda \|\ba\|_1,\quad \lambda \geq 0.
\end{equation}
Notice that \eqref{eq: lasso} is a \gls{lasso} problem that can be solved using standard optimization toolboxes (e.g., MOSEK \cite{mosek}). 
Furthermore, as suggested in \cite{becirovic2019detection}, the matrix $\mat{\Omega}$ is self-regularizing so that the regularization term $\lambda\norm{\ba}_1$ can be removed. 
The problem then reduces to a \gls{nnls} problem, which is solved by the Lawson-Hanson algorithm \cite{lawson1995solving} in \cite{becirovic2019detection}.

\subsection{Pilot Hopping in Covariance-Based Activity Detection}

It is straightforward to combine the covariance-based approach with pilot hopping. 
To do this, we ignore the correlation between different channel sub-blocks and keep only the diagonal blocks in the channel covariance matrix, i.e., 
\begin{equation}
\label{eq: block-diagonal covariance}
	\bR\approx \bdiag(\bR^{(1)},\cdots,\bR^{(P)}),
\end{equation}
where $\bR^{(p)}$ is the channel covariance within the $p$th sub-block that is assumed to be known and to have the Gramian factorization $\bR^{(p)}=\bG^{(p)}(\bG^{(p)})^\herm$. 
A motivation for ignoring the inter-block correlation is to avoid overfitting the learned low-dimensional structure to the training data, which might be generated from an oversimplified channel model that does not generalize well to real-world radio environments. 
To see this, notice that one can always learn a perfect two-dimensional representation of the standard two-ray model in \cite[Fig. 2.1]{marzetta2016fundamentals} on an infinitely large time-frequency block, while the learned representation is most probably not useful in practice.

By utilizing the (assumed) block-diagonal structure of $\bR$, the signal covariance can be expressed as 
\begin{equation}
	\bSigma_{\bgamma} = \bdiag\big(\bSigma^{(1)}_{\bgamma},\cdots,\bSigma^{(P)}_{\bgamma}\big),
\end{equation}
where $\bSigma^{(p)}_{\bgamma}\defeq \sum_{k\in[K]}\gamma_k \bS_k^{(p)}(\bS_k^{(p)})^\herm$ with $\bS_k^{(p)} = \bphi_k^{(p)}\bG^{(p)}$. Here, $\bphi_k^{(p)}$ is the $p$th sub-block in the pilot sequence of user $k$, i.e., the $k$th column of $\bPhi^{(p)}$ in \eqref{eq: pilot with hopping}.
The \gls{ml} estimation of $\bgamma$ can then be formulated as
\begin{equation}
\label{eq: f-tilde}
	\min_{\bgamma\in\Gamma}~ \widetilde{f}(\bgamma) \defeq \sum_{p=1}^P f^{(p)}(\bgamma),
\end{equation}
\begin{flalign}
\label{eq: ML objective per block}
	\mbox{where} && f^{(p)}(\bgamma) \defeq \log\big\lvert\bSigma^{(p)}_{\bgamma}\big\rvert + \tr\Big(\big(\bSigma^{(p)}_{\bgamma} \big)^{-1}\widehat{\bSigma}^{(p)} \Big) &&
\end{flalign}
with $\widehat{\bSigma}^{(p)}$ being the diagonal block in the sample covariance matrix $\widehat{\bSigma}$ corresponding to the $p$th sub-block.

It is clear that \eqref{eq: ML objective per block} has the same form as \eqref{eq: ML objective}, and the same techniques from Section \ref{sec: covariance approach} can be applied here to obtain the cost function for updating the $k$th coordinate as
\begin{equation}
\label{eq: cost 1d with pilot hopping}
	\widetilde{\cL}_k(d) \defeq \sum_{p\in\cP_k}\sum_{n\in[N]}\!\!\left(\log(1 + d\lambda_{kn}^{(p)}) - \frac{d\xi_{kn}^{(p)}}{1\!+\!d\lambda_{kn}^{(p)}} \right),
\end{equation}
where $\cP_k$ denotes the set of sub-blocks in which user $k$ is allocated a sub-pilot, i.e., $\cP_k = \{p:z_k^{(p)}\neq 0\}$. $\{\lambda_{kn}^{(p)}\}$ and $\{\xi_{kn}^{(p)}\}$ are calculated similarly as $\{\lambda_{kn}\}$ and $\{\xi_{kn}\}$ in \eqref{eq: cost 1d} for each channel sub-block. 
Since \eqref{eq: cost 1d with pilot hopping} has the same structure as \eqref{eq: cost 1d}, it can be optimized in the same way - either comparing the roots of its first-order derivative and the boundary points or performing a one-dimensional search.

\begin{algorithm}[t]
	\caption{Activity Detection with Pilot Hopping}
	\begin{algorithmic}[1]
		\label{alg: activity detection with pilot hopping}
		\REQUIRE sample covariances $\{\widehat{\bSigma}^{(p)}\}$, effective pilots $\{\bS_k^{(p)}\}$, and hopping patterns $\{\cP_k\}$
		% \REQUIRE $\{\widehat{\bSigma}^{(p)}\}$, $\{\bS_k^{(p)}\}$, and $\{\cP_k\}$
		\INITIALIZE $\gamma_k \gets 0, \forall k$ and $(\bSigma_{\bgamma}^{(p)})^{-1} \gets \sigma^{-2}\bI, \forall p$
		\FOR{$i=1,2,\cdots,I$}
		\STATE Generate $\cK$ by randomly permuting $[K]$ 
		\FOR{$k$ taken from $\cK$ by order}
		\STATE  $\forall p \in \cP_k$: $\bV^{(p)}\bD_{\blambda}^{(p)}\bV^{(p)\herm} \xleftarrow{\textup{EVD}} \bS_k^{(p)\herm}(\bSigma_{\bgamma}^{(p)})^{-1}\bS_k^{(p)} $ \label{CD-hopping step: EVD}
		\STATE $\forall p \in \cP_k$: $\widetilde{\bV}^{(p)} \gets (\bSigma_{\bgamma}^{(p)})^{-1}\bS_k^{(p)}\bV^{(p)}$ \label{CD-hopping step: V-tilde}
		\STATE $\forall p \in \cP_k, \forall n$: $\xi_n^{(p)}  \gets [\widetilde{\bV}^{(p)\herm}\widehat{\bSigma}^{(p)}\widetilde{\bV}^{(p)}]_{n,n}$ \label{CD-hopping step: xi}
		\STATE $d^* = \argmin \widetilde{\cL}_k(d)$ for $d \in [-\gamma_k,\gamma_{\max}-\gamma_k]$ \label{CD-hopping step: 1d minimization}
		\STATE $\bSigma_{\bgamma}^{(p)-1} \gets \bSigma_{\bgamma}^{(p)-1} - d^*\widetilde{\bV}^{(p)}(\mat{I} + d^*\bD_{\blambda}^{(p)})^{-1}\widetilde{\bV}^{(p)\herm}$ \label{CD-hopping step: update covariance} 
		\STATE $\gamma_k \gets \gamma_k + d^*$
		\ENDFOR
		\ENDFOR
		\ENSURE activity estimate $\bgamma = [\gamma_1,\cdots,\gamma_K]^\transp$
	\end{algorithmic}
\end{algorithm}

The detection procedure is summarized in Algorithm \ref{alg: activity detection with pilot hopping}. 

\textbf{Complexity:}
The matrix multiplications in steps \ref{CD step: EVD}, \ref{CD-hopping step: V-tilde},  \ref{CD-hopping step: xi}, and \ref{CD-hopping step: update covariance} have complexity $\cO(|\cP_k|\tau^2N)$. 
The \gls{evd} in step \ref{CD-hopping step: EVD} has complexity $\cO(|\cP_k|N^3)$.
However, as the derivative of $\widetilde{\cL}_k(d)$ in \eqref{eq: cost 1d with pilot hopping} is a polynomial of degree $2\lvert\cP_k\rvert N-1$, finding the root requires $\cO(\lvert \cP_k\rvert^3 N^3)$ arithmetic operations.
In case that $\abs{\cP_k} N$ is large, employing a one-dimensional search can be more efficient, as the complexity does not scale with $\abs{\cP_k}$ nor $N$.
The overall complexity of Algorithm \ref{alg: activity detection with pilot hopping} is $\cO(I K \abs{\cP_k} N (\tau^2 + N^2))$. 
Notice that Algorithm \ref{alg: activity detection with pilot hopping} can be parallelized when the users can be partitioned into groups that transmit in disjoint sub-blocks. For example, when each user transmits in only one sub-block, i.e., $|\cP_k| = 1, \forall k$, the \gls{bs} can perform activity detection in each sub-block separately.

\subsection{Hopping Pattern Design}

Another critical design aspect in pilot hopping-based activity detection is the selection of hopping patterns for each user. 
In \cite{becirovic2019detection}, the hopping patterns are selected uniformly at random, where the patterns $\{z_k^{(p)}\}$ are generated in an \gls{iid} manner with uniform distribution across $[J]$. 
We modify this method to restrict each user to selecting only  $D$ sub-blocks, i.e., $|\cP_k| = D, \forall k$, to control the average overload factor $DK/(P\tau)$ in each sub-block. The method in \cite{becirovic2019detection} can be seen as a special case when $D=P$. 

\begin{figure}
	\centering
	\includegraphics[width=\linewidth]{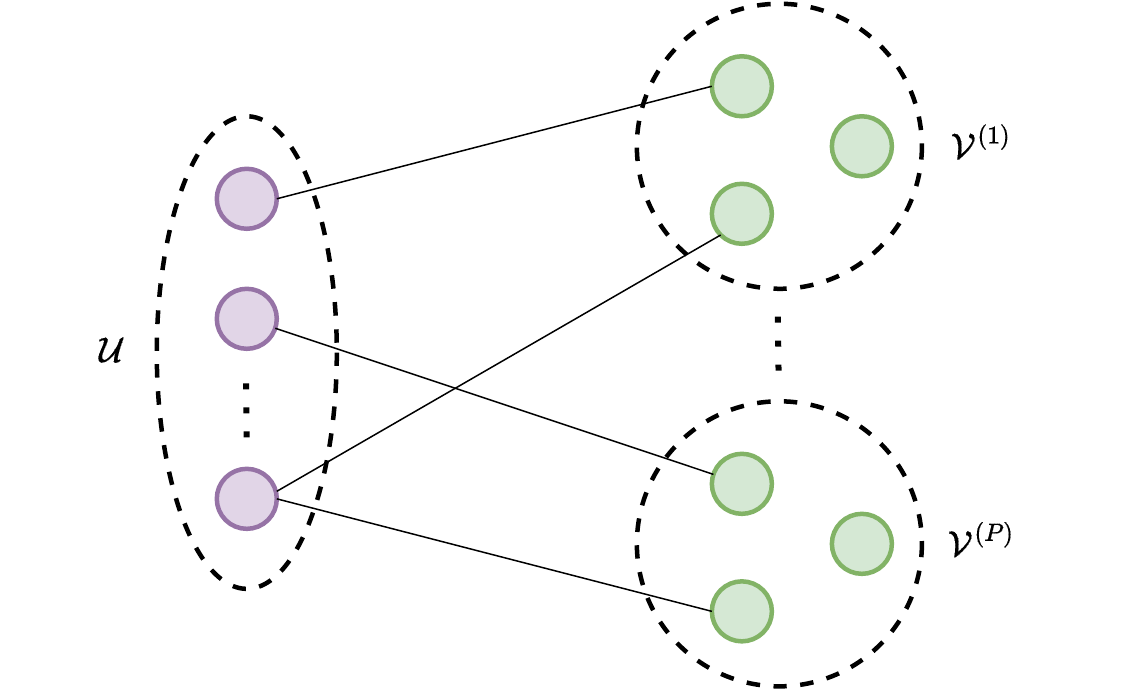}
	\caption{Pilot hopping system as a bipartite graph.}
	\label{fig: configuration-model}
\end{figure}

In the random pilot hopping generation scheme, however, the number of users selecting different sub-blocks, as well as sub-pilots within each sub-block, can vary significantly.
To address this imbalance in the hopping system, we propose a new hopping pattern generation scheme, inspired by the configuration model in random graph generation \cite[Ch. 5.3]{latora2017complex}. 
Specifically, the system can be seen as a bipartite graph $\cG=(\cU,\cV,\cE)$, as shown in Fig. \ref{fig: configuration-model}, where the first vertex set $\cU$ denotes the users, the second vertex set $\cV\defeq\cup_{p\in[P]}\cV^{(p)}$ represents sub-pilots with $\cV^{(p)}$ being the sub-pilots in the $p$th sub-block.
We aim to generate the edge set $\cE\defeq \cup_{p\in[P]}\cE^{(p)}$ with $\cE^{(p)}$ representing the edges between $\cV^{(p)}$ and $\cU$ with all sub-pilot vertices in $\cV$ having nearly the same degrees, differing at most by one, i.e., $\lfloor DK/(PJ)\rfloor \leq\deg(v) \leq \lfloor DK/(PJ)\rfloor + 1, \forall v\in\cV$, all sub-blocks having nearly the same number of edges, i.e., $\lfloor DK/P\rfloor \leq |\cE^{(p)}| \leq \lfloor DK/P\rfloor + 1, \forall p\in[P]$, and each user vertex in $\cU$ having degree $D$ with at most one edge to each $\cV^{(p)}$.
The procedure for generating $\{\cE^{(p)}\}$ is summarized in Algorithm \ref{alg: configuration model}. 
The algorithm iterates over a randomly permuted list of users and randomly chooses for each user $D$ sub-blocks with the minimal number of edges and one sub-pilot with the lowest degree in each selected sub-block to form a new edge.
The conversion from the edge lists $\{\cE^{(p)}\}$ to the hopping patterns $\{z_k^{(p)}\}$ is straightforward: if there is an edge connecting the vertex corresponding to user $k$ and the vertex for the $j$th sub-pilot in the $p$th sub-block, we set $z_k^{(p)}=j$. If no vertex in $\mathcal{V}^{(p)}$ is connected to the vertex representing user $k$, we assign $z_k^{(p)}=0$.

\begin{algorithm}[t]
	\caption{Hopping Pattern Generation}
	\begin{algorithmic}[1]
		\label{alg: configuration model}
		\REQUIRE user vertices $\cU$, sub-pilot vertices $\{\cV^{(p)}\}$, and $D$
		\INITIALIZE $\cE^{(p)} \leftarrow \emptyset, \forall p\in[P]$
		\FOR{$u\in\cU$}
		\STATE $\widetilde{\cP}\leftarrow\emptyset, i \leftarrow 0$
		\WHILE{$|\widetilde{\cP}| < D$}
		\STATE $\widetilde{\cP} \leftarrow \{p: |\cE^{(p)}| \leq \min_{p'\in[P]}|\cE^{(p')}| + i\}$
		\STATE $i \leftarrow i + 1$
		\ENDWHILE
		\WHILE{$\deg(u) < D$}
		\STATE choose $p\in\widetilde{\cP}$ randomly and remove $p$ from $\widetilde{\cP}$
		\STATE $\widetilde{\cV}^{(p)} \leftarrow \argmin_{v\in\cV^{(p)}} \deg(v)$
		\STATE choose $v\in\widetilde{\cV}^{(p)}$ randomly and add $(u,v)$ to $\cE^{(p)}$ 
		\ENDWHILE
		\ENDFOR
		\ENSURE $\{\cE^{(p)}\}$
	\end{algorithmic}
\end{algorithm}

\subsection{The Complete Framework}
\label{sec: complete framework}

\begin{figure}
	\centering
	\includegraphics[width=0.9\linewidth]{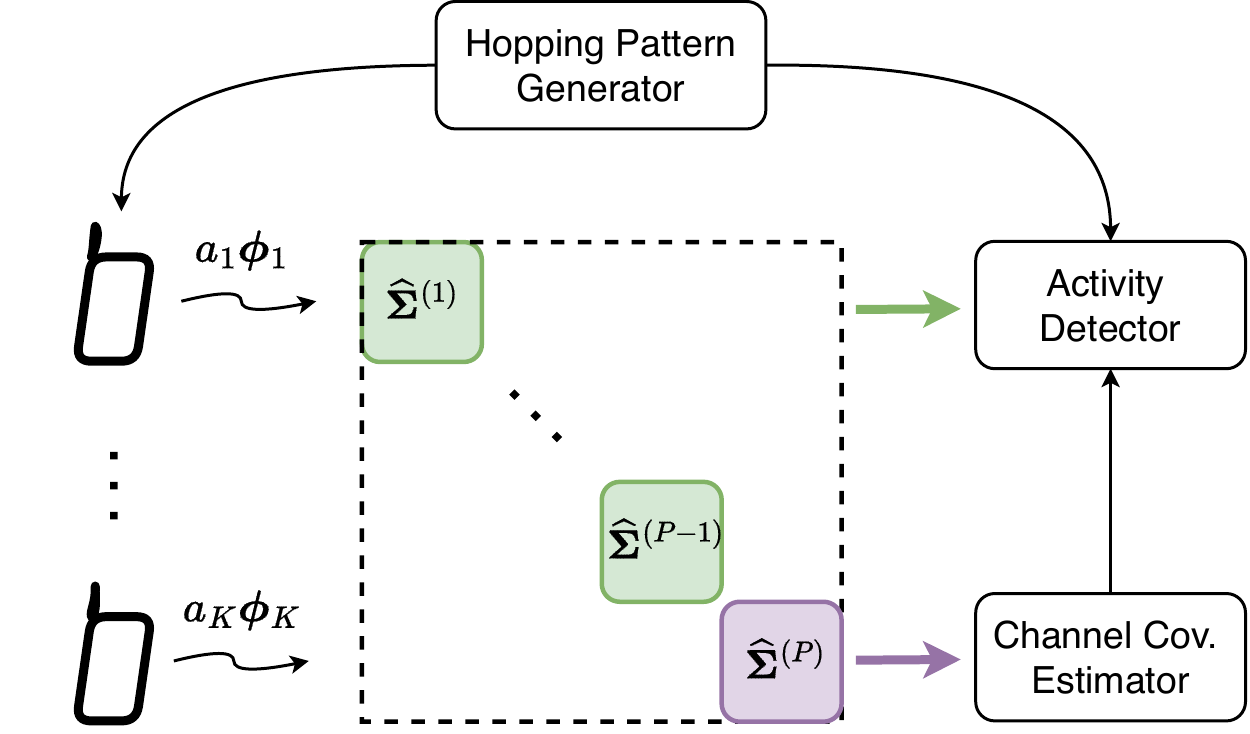}
	\caption{The complete activity detection framework.}
	\label{fig: complete-framework}
\end{figure}

We can combine the covariance estimation using dedicated, all-one pilots with the pilot hopping to reduce the overhead.
We assume the channel is nearly \gls{wssus} so that the diagonal blocks in the channel covariance $\bR$ are approximately identical, i.e., $\bR^{(1)}\approx \cdots \approx \bR^{(P)} \defeq \bR_0$. Then, it becomes sufficient to use only one sub-block for covariance estimation, leaving the remaining ones for activity detection.
Our detection framework, as shown in Fig. \ref{fig: complete-framework}, is as follows: 
All active users send the all-one sequence in the $P$th sub-block for the \gls{bs} to estimate the channel covariance.
In the remaining $P-1$ channel sub-blocks, these active users follow their unique hopping patterns for pilot transmission.
The \gls{bs} performs activity detection using the received pilot signals as well as the bases obtained from the covariance estimate.

\begin{remark}
\label{remark: sub-block structure}
	The division of a channel block into multiple sub-blocks, and especially the size of these sub-blocks, need to be carefully designed. 
	For strongly correlated fading, we should use larger sub-blocks to increase the gain from coherent processing of the samples.  
	Conversely, in scenarios with less correlated fading, forming smaller sub-blocks helps exploit diversity. 
	There is also a tradeoff between computational complexity and detection performance.
	From a data-driven perspective, we could first estimate the channel covariance matrix $\bR$ and calculate the eigenvalues of its sub-blocks of different sizes to determine the sub-block size.
	We should select the sub-block size such that the number of dominant eigenvalues of the sub-blocks is relatively small, allowing for a good low-dimensional approximation.
	One numerical example is provided in Section \ref{subsec: visualization of channel}, where we show that the selected sub-block size gives an accurate order-3 approximation of the channel covariance matrix.
	Alternatively, if the channel can be well approximated by an existing model, for example as in \cite[Ch. 2]{stuber2001principles}, and the relevant parameters are known, we can choose the sub-block size based on the channel auto-correlation functions.
	In this paper we focus on the case where the channel is nearly \gls{wssus} and the channel covariance needs to be frequently re-estimated.
	In case that the channel is not \gls{wssus} but the channel covariance changes slowly so that re-estimation is infrequent, we can use more complicated strategies.
	One idea is to use the magnitudes of the elements in the channel covariance as edge weights to form a graph and apply graph partitioning methods (see, for example, \cite[Ch. 9]{latora2017complex}) to determine the sub-block structure -- in this way, the elements in the same sub-block are not necessarily adjacent to each other.
	The design of data-driven approaches to determine the optimal sub-block structure could be an interesting direction for future research.

\end{remark}

\section{Numerical Results}

\label{sec: simulations}

\subsection{Generation of Channels}

\label{sec: generation of channel}

We first generate the \glspl{cir} by using the improved sum-of-sinusoids method in \cite{xiao2006novel}. Since the same procedure will be repeated for each user-antenna pair, we ignore the subscript ``$km$''. For a channel with $\Npath$ paths at different delays, we generate each path independently, and the time-varying (complex) amplitude of the $i$th path is given by
\begin{equation}
	q_i(t) = \frac{1}{\sqrt{\Nsin}} \sum_{n\in[\Nsin]} e^{j(\omega_d t \cos\alpha_n + \psi_n)}
\end{equation}
with 
$
	\alpha_n = \frac{2\pi n + \zeta_n}{\Nsin}, n\in[\Nsin],
$
where $\Nsin$ is the number of sinusoids, $\omega_d$ is the maximum Doppler frequency in radians, $\psi_n$ and $\zeta_n$ are i.i.d. distributed over $[-\pi,\pi)$.

\begin{figure}
	\centering
	\includegraphics[width=5cm]{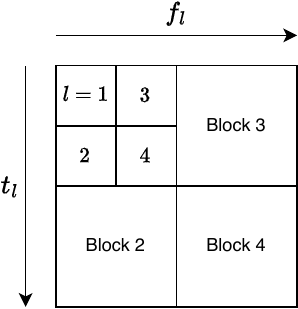}\hspace{1cm}
	\caption{The mapping from $(t_l,f_l)$ to $l$ for $T=F=4$ and $P_\textup{time} = P_\textup{freq} = 2$.}
	\label{fig: channel-map}
\end{figure}

We denote the sampling rate by $B$ which is identical to the system bandwidth, and the impulse response of the pulse shaping filter as $p(\cdot)$.
The discrete-time impulse response of the multipath-fading channel is given by
\begin{equation}
	q_{t_l \ell} = \sum_{i\in[\Npath]} \sqrt{c_i} q_i\left(\frac{t_l-1}{B}\right) p\left(\frac{\ell-\tau_i}{B}\right),
\end{equation}
where $t_l\in[T]$, the fractional power $c_i$, and the delay $\tau_i$ of different paths are determined by a power delay profile defined in, for example, \cite{3gppTR1,3gppTR2}. Here, $\ell$ is the time-lag index, which is an integer ranging from $\ell_{\min}$ to $\ell_{\max}$. The smallest time-lag $\ell_{\min}$ is a negative integer, e.g., $-6$, and $\ell_{\max}=\lceil B\tau_{\textup{exc}} \rceil - \ell_{\min}$ with $\tau_{\textup{exc}}=\max_{i\in[\Npath]}\{\tau_i\}$ being the maximum excess delay. For each $t_l$, we apply the \gls{dft} to $\bx_{t_l}=[q_{t_l\ell_{\min}},\cdots,q_{t_l\ell_{\max}}]^\transp$ to obtain the frequency response at $f_l\in[F]$ as
\begin{equation}
	Q_{t_lf_l} = \sum_{i\in[\ell_{\max}-\ell_{\min}+1]} [\vect{x}_{t_l}]_i e^{-j\frac{2\pi (i-1) (f_l-1)}{\Nsub}},
\end{equation}
where $\Nsub \geq F$ is the total number of sub-carriers in this \gls{ofdm} system (we use the first $F$ sub-carriers for pilot transmission).
These coefficients are arranged into the length-$L$ channel vector $\bh = [h_1,\cdots,h_L]^\transp$ where $h_l$ equals to $Q_{t_lf_l}$ with an invertible mapping from $(t_l,f_l)$ to $l$.
For simplicity, we assume that the total number of sub-blocks can be factorized as $P=\Ptime\Pfreq$, where $\Ptime$ is an integer that divides the number of OFDM symbols $T$ and $\Pfreq$ divides the number of sub-carriers $F$.
We use a mapping between $(t_l,f_l)$ and $l$ as illustrated in Fig. \ref{fig: channel-map}.

\subsection{System Setup}

\label{subsec: system setup}

We consider a single-cell network with $K=4000$ users. 
Each user becomes active independently with a probability of $\epsilon=0.1$.
The \gls{bs} has $M=100$ receive antennas.
We consider a peak power constraint and ideal channel inversion power control so that the \gls{snr} equals 0 dB for all users, i.e., $\beta_1/\sigma^2 = \cdots = \beta_K/\sigma^2 = 1$.
The channels are generated using the procedure in Section \ref{sec: generation of channel} with $T=12$ and $F=36$, resulting in a channel dimension of $L=432$.
The carrier frequency is set to 30 GHz which falls into the frequency range 2  in 5G New Radio.
The sub-carrier spacing is set to 30 kHz \cite{lin20195g}. 
The \gls{dft} size is set to $128$. 
The number of sinusoids in the channel simulator is $\Nsin=20$. 
For pulse shaping, we use a \gls{rrc} filter with a roll-off factor of $0.22$.
In each Monte-Carlo trial, the channel model is randomly selected from TDL-A, TDL-B, and TDL-C in \cite[Sec. 7.7.2]{3gppTR2} and \gls{tux}, \gls{rax}, and \gls{htx} in \cite[Sec. 5]{3gppTR1}, where the delays are randomly scaled to have a \gls{rms} delay spread between 0.5 and 1.5 microseconds.
The mobile speed is randomly selected between 80 and 160 km/h.
The time-frequency grid is partitioned into $P=9$ channel sub-blocks with $\Ptime=\Pfreq=3 $.
Unless otherwise stated, we use $D=1$, $J=4000$, and Algorithm \ref{alg: configuration model} for hopping pattern generation. 
The sub-pilots are randomly generated from complex Gaussian and are normalized to have unit average energy per symbol.
The performance of the detection algorithms is evaluated using \gls{roc} curves, each one generated using 3000 independent Monte Carlo trials.

\subsection{Is Pilot Hopping Useful? -- A Case Study}

\begin{figure}
	\begin{subfigure}{0.5\textwidth}
		\centering
		\includegraphics[width=\textwidth]{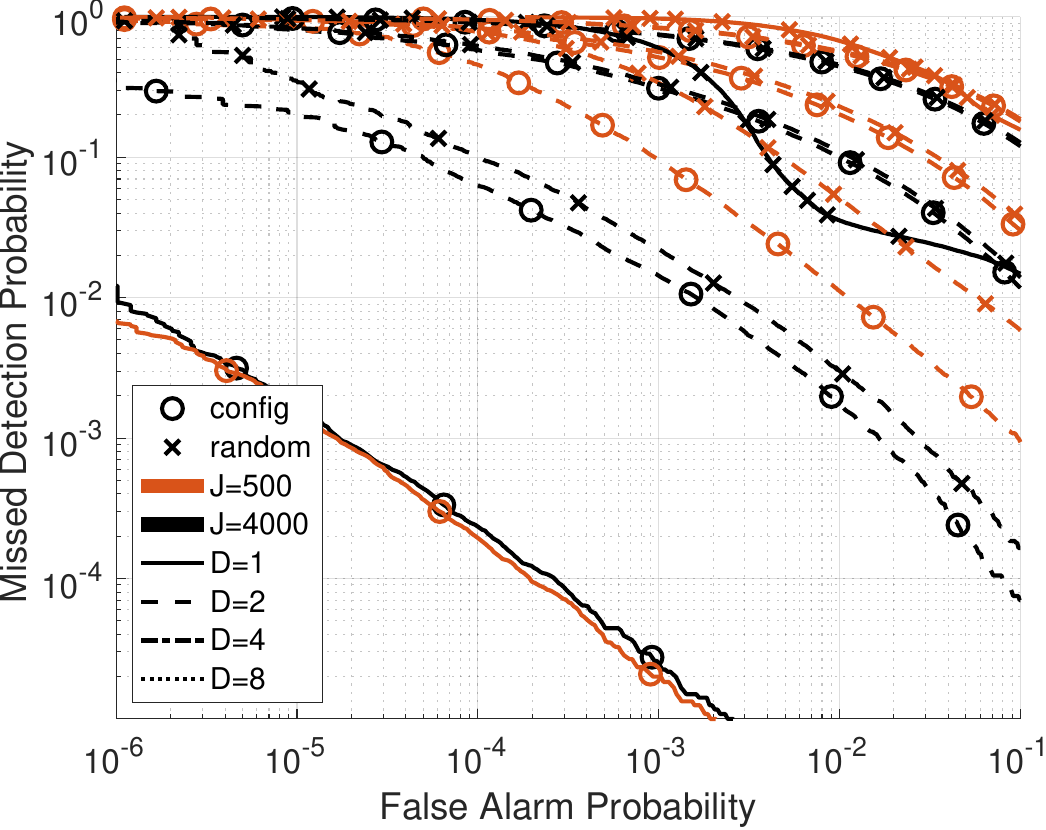}
		\caption{without flashlight interference}
		\vspace{0.3cm}
		\label{result: hopping-nointerf}
	\end{subfigure}
	\begin{subfigure}{0.5\textwidth}
		\centering
		\includegraphics[width=\textwidth]{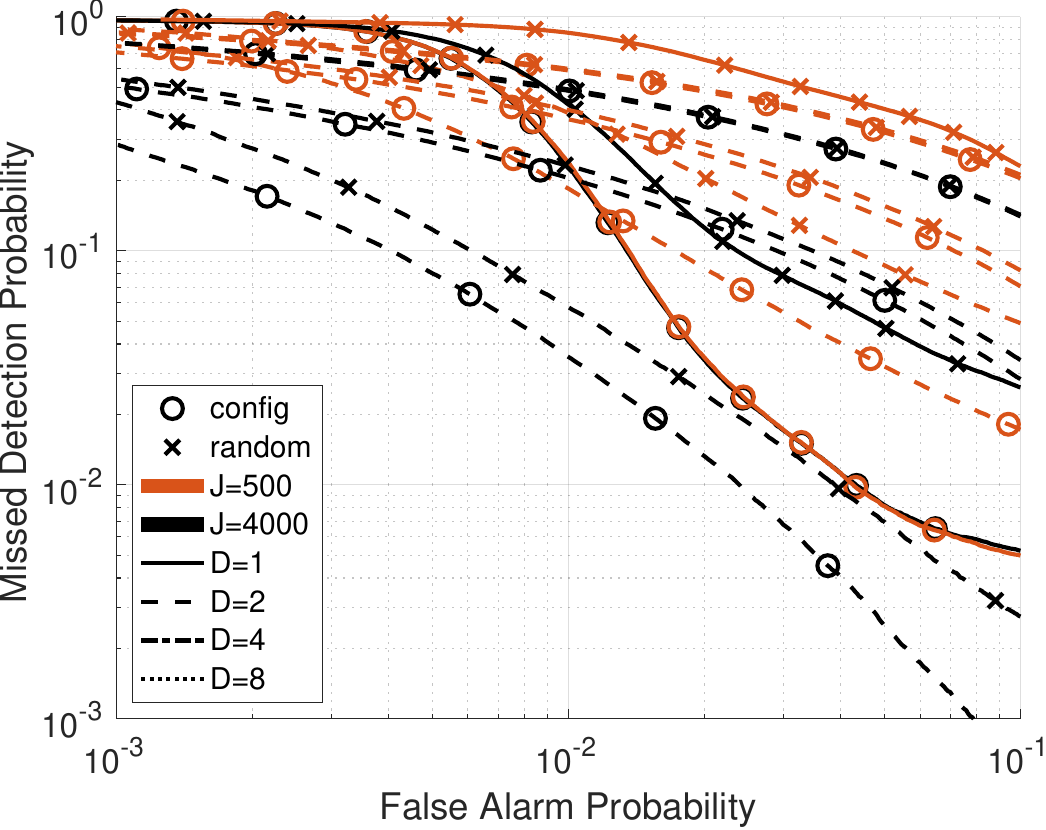}
		\caption{with flashlight interference}
	\label{result: hopping-interf}
	\end{subfigure}
	\caption{Detection performance with different hopping schemes.}
	\label{result: hopping}
\end{figure}

The \gls{roc} curves of different pilot hopping schemes are presented in Fig. \ref{result: hopping-nointerf}. 
We observe that the hopping patterns generated by the configuration model, as described in Algorithm \ref{alg: configuration model}, significantly outperform the random patterns.

{Choice of $D$:}
When using good hopping patterns, it does not appear advantageous to let users make multiple transmissions in different channel sub-blocks.
Instead, the best detection performance is achieved when each active user transmits in only $D=1$ sub-block, which suggests that one should simply partition the users into disjoint groups and allocate each group a dedicated channel sub-block -- this scheme more closely resembles \emph{scheduling} than hopping. 
One interpretation of this result is that when sufficient spatial diversity has been achieved by a large antenna array, it becomes unnecessary to explore extra time or frequency diversity at the cost of increased interference.  
The concurrent work \cite{dejesus2024assessment} reveals a similar observation: one should not trade sparsity for diversity in scenarios with limited radio resources.

{Choice of $J$:}
Since our approach combines pilot hopping with covariance-based activity detection, using orthogonal sub-pilots ($J=\tau$) with collision but no contamination, and unique sub-pilots ($J=K$) with no collision but severe contamination,  respectively, it is of interest to explore the  tradeoff between these two design principles.
However, as suggested by our results, users should not reuse sub-pilots, and better performance can be achieved by assigning a unique sub-pilot to each user within every channel sub-block. (In our setup, users have unique sub-pilots when $J\geq DK/P=500D$.)

While the findings so far might suggest that pilot hopping is not useful, this conclusion could vary based on the scenario. 
For instance, in real-world situations, the receiver might experience ``flashlight interference'' from neighboring cells (due to imperfect coordination/scheduling in multi-cell networks; see \cite{khosravirad2022communications}) or encounter accidental blocking effects, leading to signal contamination or weakening in certain channel sub-blocks. 
Under such circumstances, exploiting additional diversity through pilot hopping becomes important. 
Specifically, when a user transmits its pilot in only one of the sub-blocks, there is a possibility that its signal will be contaminated by this interference, leading to a higher probability of detection error. 
Conversely, if users utilize multiple sub-blocks to exploit diversity, even though their signals may be contaminated in one sub-block, the BS can still leverage the diversity in the other sub-blocks to effectively detect the users.
In Fig. \ref{result: hopping-interf}, we recalculate the results from Fig. \ref{result: hopping-nointerf}, introducing 100 interfering users who transmit randomly generated complex Gaussian signals within a randomly selected channel sub-block. 
The results show that enabling users to transmit pilots over multiple sub-blocks enhances robustness against flashlight interference.

\subsection{Comparison of Different Bases}

In addition to the learned bases (referred to as the \gls{pca} bases henceforth), we consider three other schemes as baselines:\footnote{The \gls{bwl}/\gls{dft} bases were used in \cite{jiang2022massive} and \cite{jiang2022statistical}, respectively, for frequency selectivity. We extend their methods by incorporating time variations.   }
(We use block size $T\times F$ for illustration, which will be adjusted accordingly for pilot hopping. We use $N=3$ unless otherwise stated.)
\begin{itemize}
	\item Block-fading: the channel is assumed to be constant within each sub-block, i.e., $\bG = \bone_L$ and $N=1$.

	\item \gls{bwl} bases: the channel varies linearly over time and frequency. To be specific, consider two vectors
	\begin{align*}
		\bu_T \propto&~ [-1,-1+2/(T-1),\cdots,1]^\transp\in\R^T\\
		\bu_F \propto&~ [-1,-1+2/(F-1),\cdots,1]^\transp\in\R^F
	\end{align*}
	representing the linear variations, where $\propto$ means that the vectors are normalized, i.e., $\|\bu_T\|^2 = T, \|\bu_F\|^2 = F$. 
	Let $\widetilde{\bG}=[\bone_L,\bone_F\otimes\bu_T,\bu_F\otimes\bone_T]$, where $\otimes$ represents the Kronecker product, and $\bone_L$ accounts for the average channel gain. The basis matrix is given by $\bG=\widetilde{\bG}\bD_\bx^{\frac{1}{2}}$, where $\bx\in\R^3$ denotes the variances after projecting the channel vectors onto each basis.

	\item \gls{dft} bases: We select $\bv_T$ to be one column from the $T\times T$ \gls{dft} matrix, and $\bv_F$ from the $F\times F$ \gls{dft} matrix. (We traverse all columns and choose the ones giving the best approximation.) Similar to the \gls{bwl} case, we choose $\bG = \widetilde{\bG}\bD_\bx^{\frac{1}{2}}$ with $\widetilde{\bG} = [\bone_L,\bone_F\otimes\bv_T,\bv_F\otimes\bone_T]$.
\end{itemize}

Similar to the learning of the low-dimensional structure in Section \ref{sec: cov estimation}, to determine the vector $\bx$ in the \gls{bwl}/\gls{dft} bases, we can form a target matrix $\bUpsilon \defeq \widehat{\bSigma}_0 - \sigma^2\bI$, where $\widehat{\bSigma}_0$ is the sample covariance after sending all-one pilots, and set $\bx = \diag(\widetilde{\bG}^\herm\bUpsilon\widetilde{\bG})$ which minimizes $\|\bUpsilon - \widetilde{\bG}\bD_\bx\widetilde{\bG}^\herm\|_\sfF$.

\begin{figure}
	\centering
	\includegraphics[width=\linewidth]{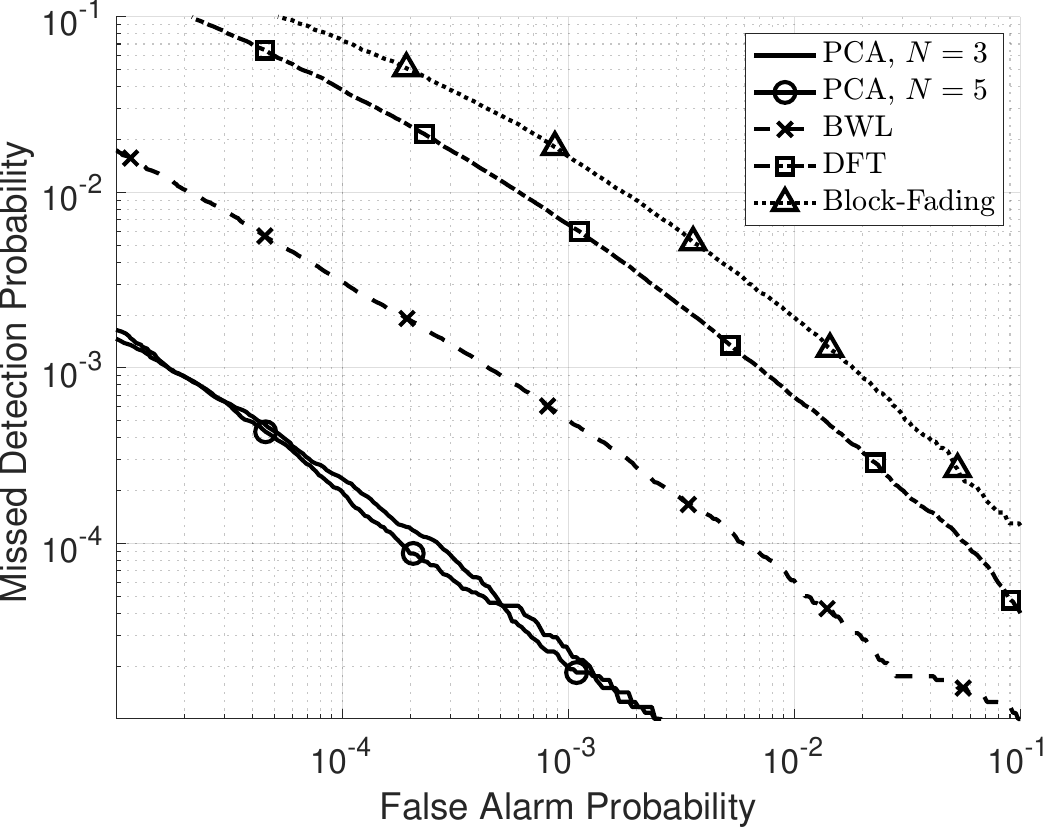}
	\caption{Performance comparison of different bases.}
	\label{result: comparison of bases}
\end{figure}

The detection performance with different bases for the low-dimensional channel approximation is shown in Fig. \ref{result: comparison of bases}. One can observe that the \gls{pca}-based bases outperform the competing models by a considerable margin.
However, using a higher approximation order ($N=5$) does not necessarily give significantly better performance than a lower approximation order ($N=3$) even at the cost of increased computational complexity. 
This is because our knowledge of the low-dimensional channel subspace is imperfect. The higher-order bases may be subject to larger estimation noise.

\subsection{Performance with Different Learning Overheads}

\begin{figure}
	\centering
	\includegraphics[width=\linewidth]{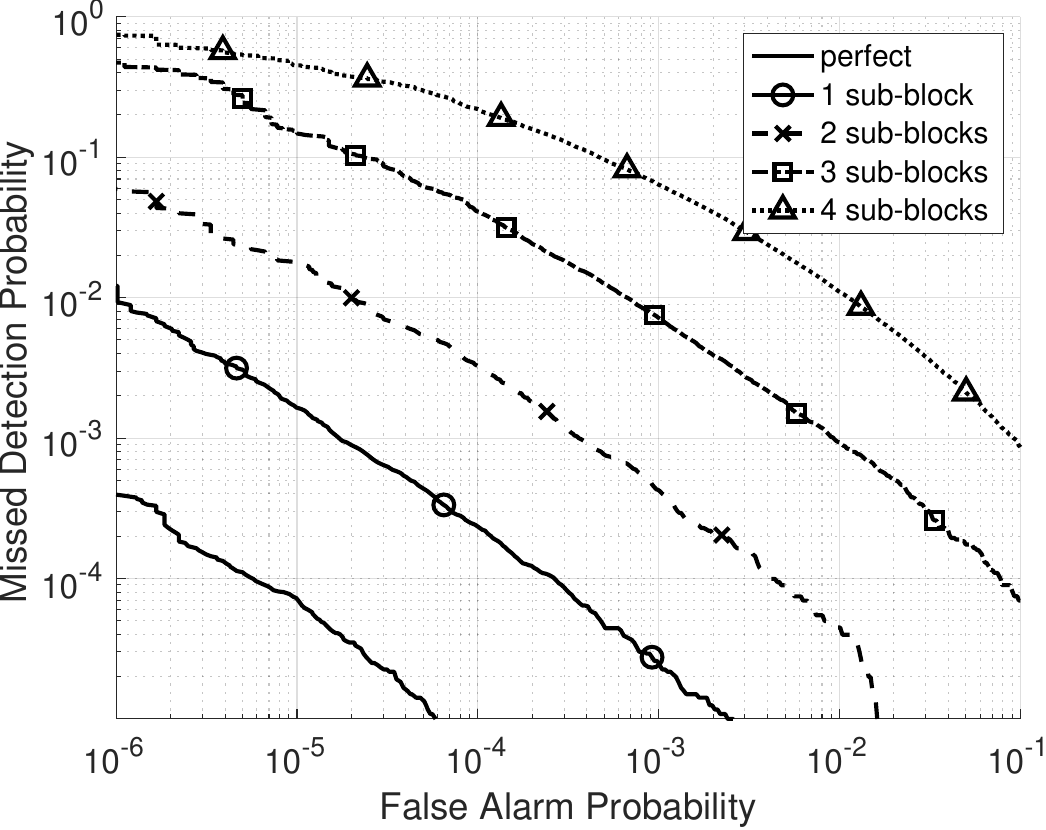}
	\caption{Detection performance with different numbers of sub-blocks dedicated for low-dimensional structure learning.}
	\label{fig: different learning overhead}
\end{figure}

It is of interest to know how much overhead one should spend on learning the low-dimensional channel structure. 
On the one hand, we want to learn the structure more accurately so that the activity detection algorithm can work better. 
On the other hand, due to the limited radio resources, we have fewer resources for activity detection as we spend more resources on structure learning. 
It is difficult to draw a general conclusion since there is no explicit objective function that can be feasibly optimized. 
Additionally, as discussed in Section \ref{sec: learn channel covariance}, the pilot design can also play an important role in this problem, and our choice of all-one pilots is only a special case.
In Fig.~\ref{fig: different learning overhead}, we compare the detection performance when choosing different numbers of dedicated sub-blocks for low-dimensional channel structure learning.
We achieve the best performance when using only one sub-block for structure learning.
This suggests that by sending all-one pilots, we can estimate the low-dimensional structure with sufficient accuracy, and there is no need to use more sub-blocks.

\begin{figure}
	\centering
	\includegraphics[width=\linewidth]{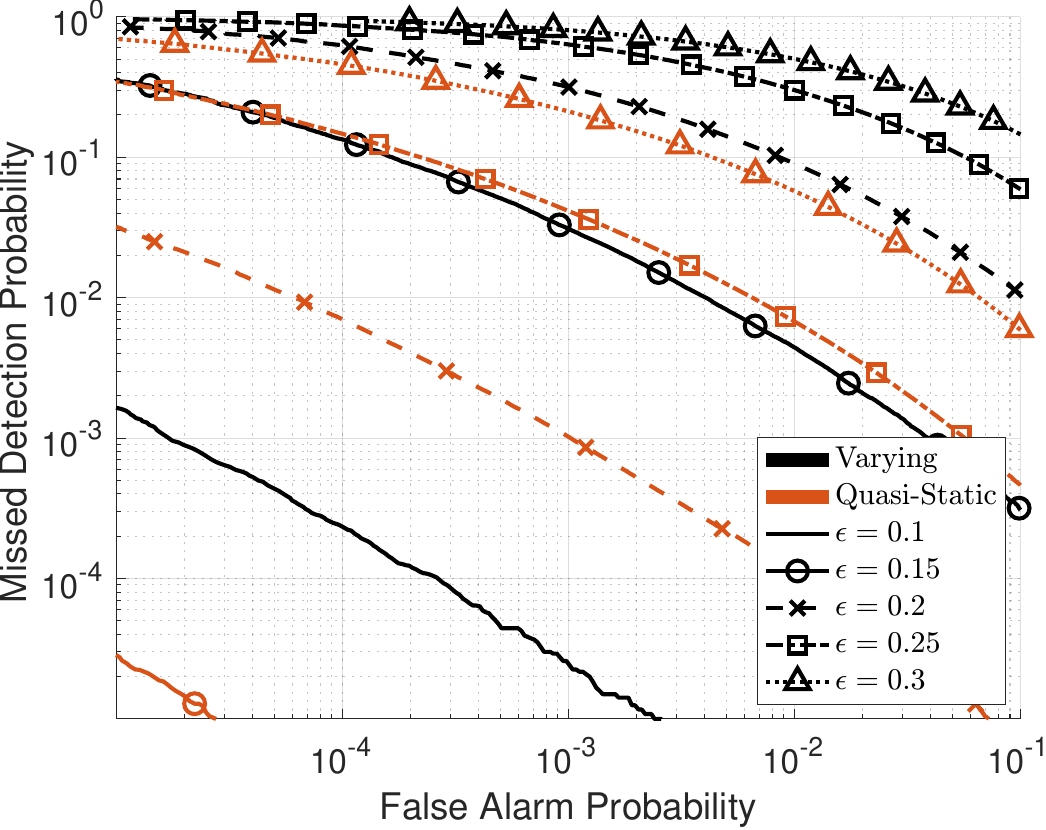}
	\caption{Performance with different active probabilities $\epsilon$.}
	\label{fig: active prob}
\end{figure}

\subsection{Performance with Different Active Probabilities}

In Fig. \ref{fig: active prob}, we can observe that the detection performance degrades rapidly when users access the channel with higher probabilities. 
Furthermore, compared with a quasi-static block-fading channel, where the channel stays constant within each sub-block, the continuously varying channels make the detection problem more challenging, as fewer users can be simultaneously supported with the same amount of radio resources.
This is unavoidable, as there is less coherence that can be exploited in the continuously varying channels, and also because additional radio resources need to be allocated to the learning of the low-dimensional channel structure.

\subsection{Runtime}
\label{sec: runtime}

Activity detection using Algorithm \ref{alg: activity detection with pilot hopping} for 10 global iterations takes around 3 seconds for the system setup in Section \ref{subsec: system setup}.
(We use a one-dimensional search to minimize \eqref{eq: cost 1d with pilot hopping} directly. The runtime was obtained on a standard PC using Matlab.)

We note that our implementation of the algorithms is by no means optimized for speed. 
There are potential improvements that could reduce runtime. 
For instance, as showcased in \cite{henriksson2020architecture}, the runtime of the coordinate decent algorithm can be significantly reduced by using properly designed computational architectures.
Additionally, the active set selection algorithm in \cite{wang2021efficient} could be used to avoid iterating over all users in each global iteration of Algorithm \ref{alg: activity detection with pilot hopping} to accelerate the coordinate descent algorithm.
Designing more computationally efficient algorithms is an important direction for future research.

\subsection{Visualization of Channels}
\label{subsec: visualization of channel}

\begin{figure}
	\centering
	\includegraphics[width=\linewidth]{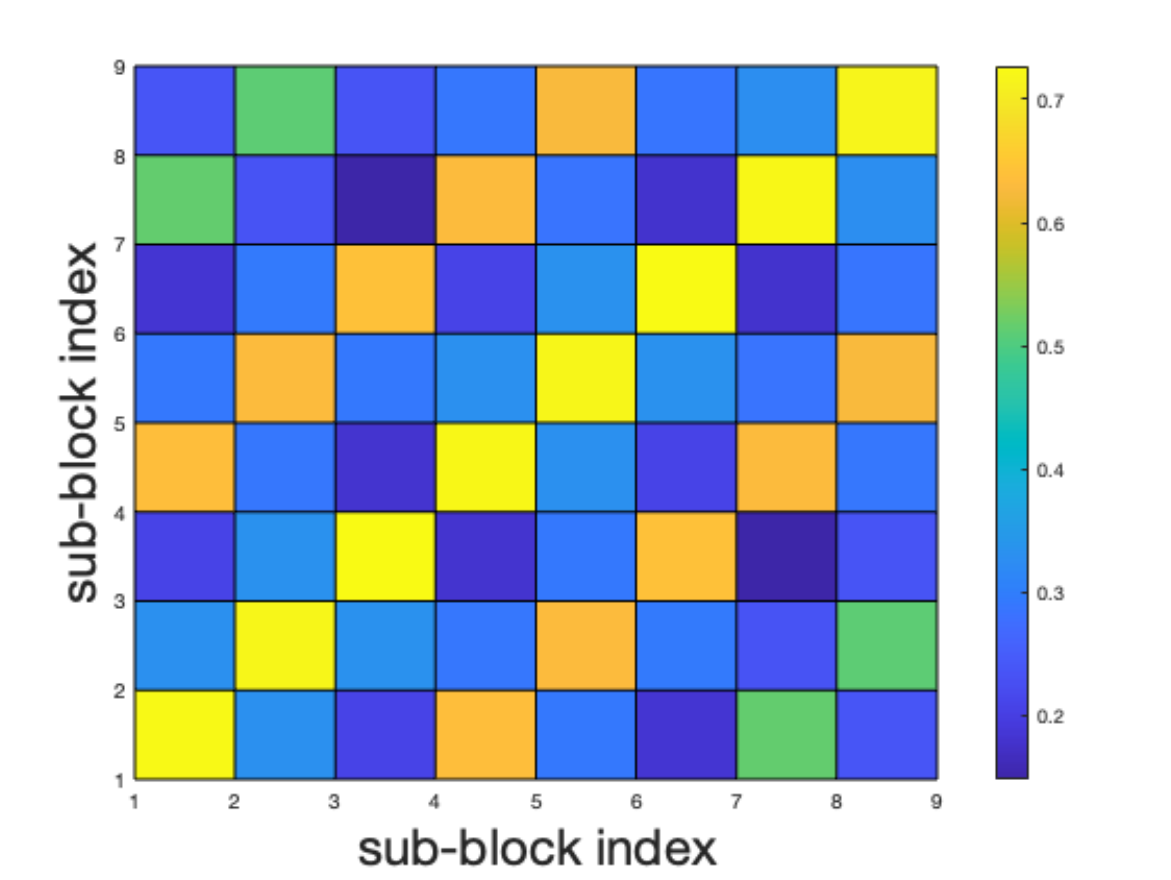}
	\caption{The magnitude of elements in the channel covariance matrix averaged over each sub-block pair.}
	\label{fig: channel-covariance}
%	\vspace{-0.5cm}
\end{figure}

\begin{figure*}
	\centering 
	\begin{subfigure}[b]{0.32\textwidth}
		\centering
		\includegraphics[width=\textwidth]{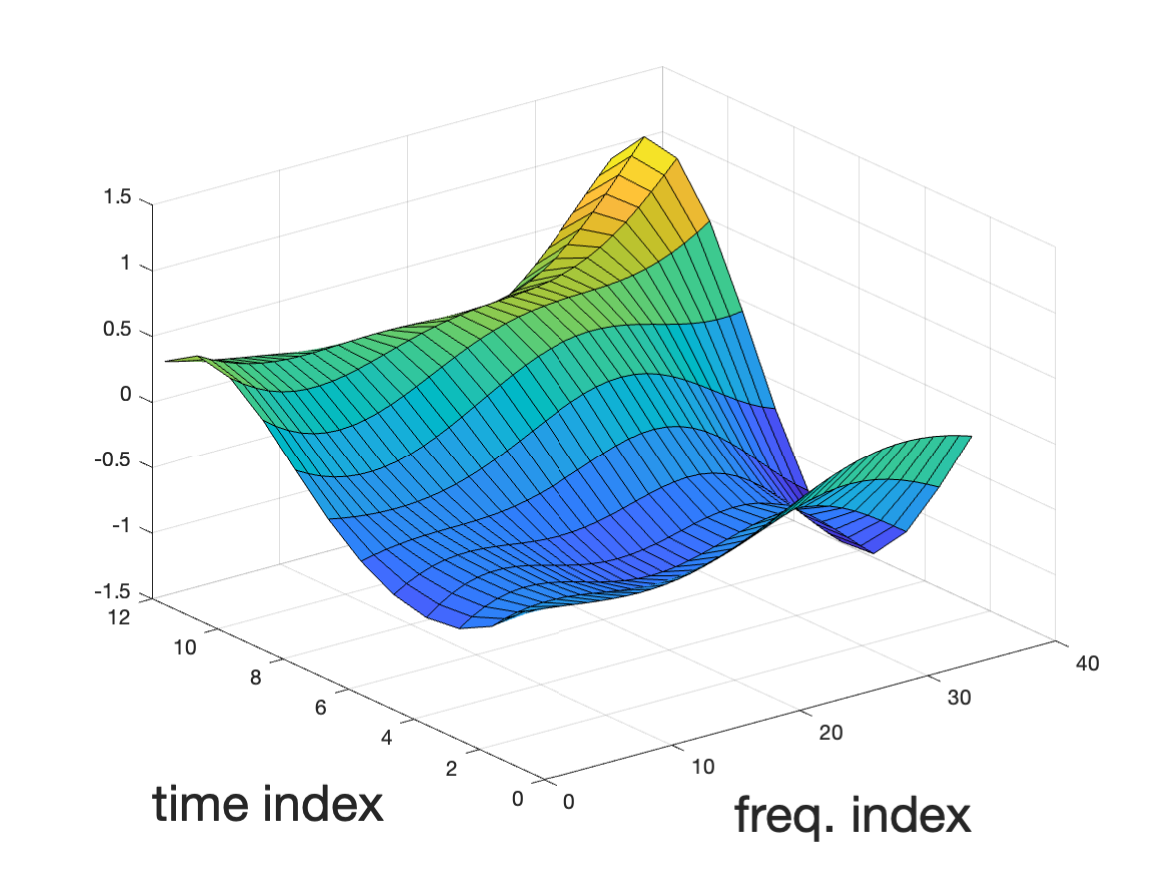}
		\caption{true channel}
	\end{subfigure}
	\hfill
	\begin{subfigure}[b]{0.32\textwidth}
		\centering
		\includegraphics[width=\textwidth]{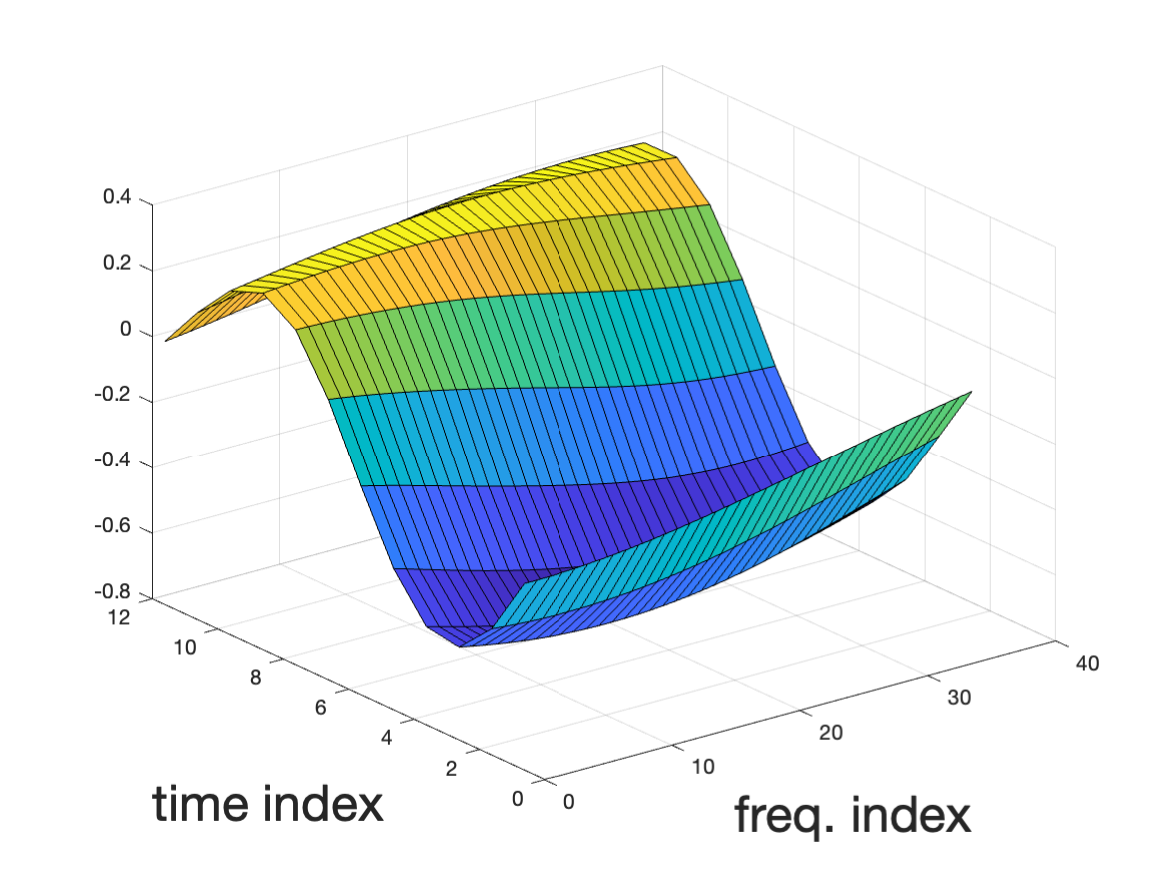}
		\caption{approx.}
	\end{subfigure}
	\hfill
	\begin{subfigure}[b]{0.32\textwidth}
		\centering
		\includegraphics[width=\textwidth]{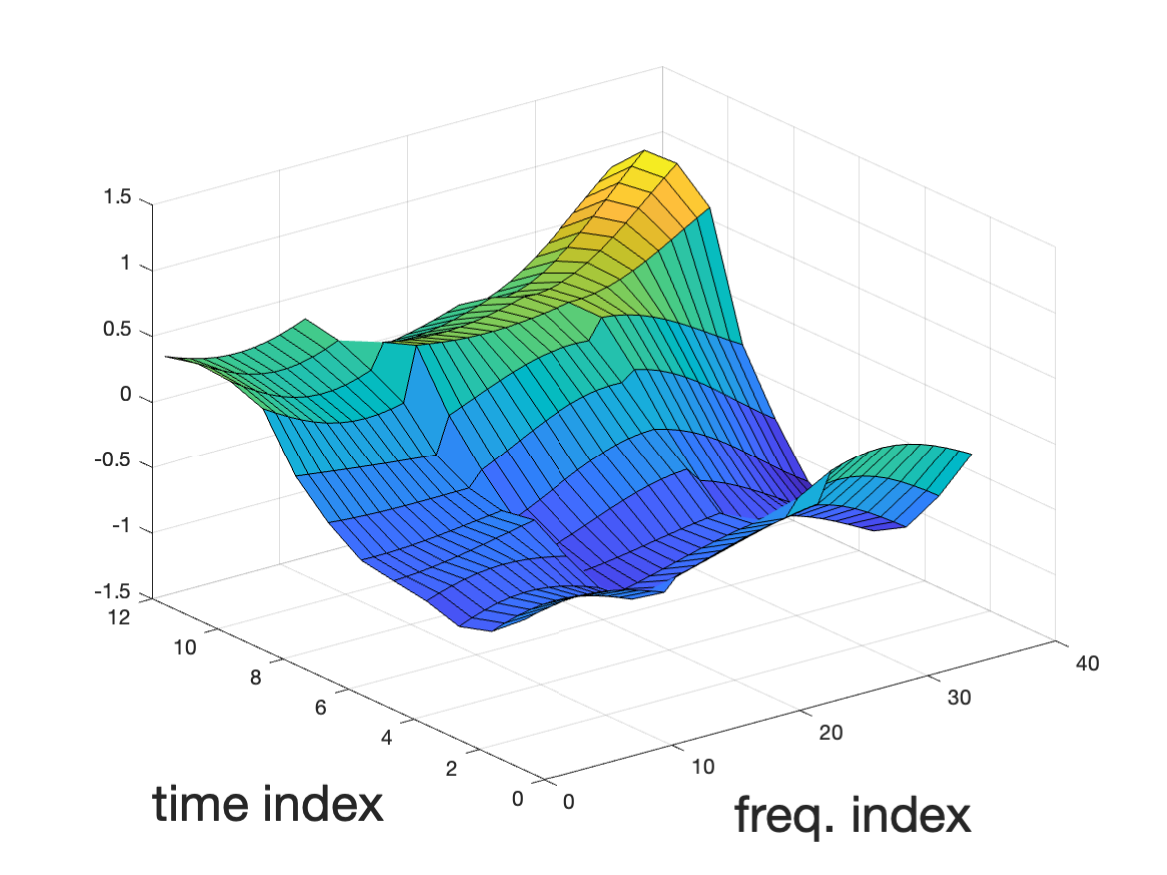}
		\caption{approx. per sub-block}
	\end{subfigure}
	\caption{Channel visualization and its order-3 approximation.}
	\label{fig: channel-blcok}
%	\vspace{-0.5cm}
\end{figure*}

\begin{figure*}
	\centering
	\begin{subfigure}[b]{0.32\textwidth}
		\centering
		\includegraphics[width=\textwidth]{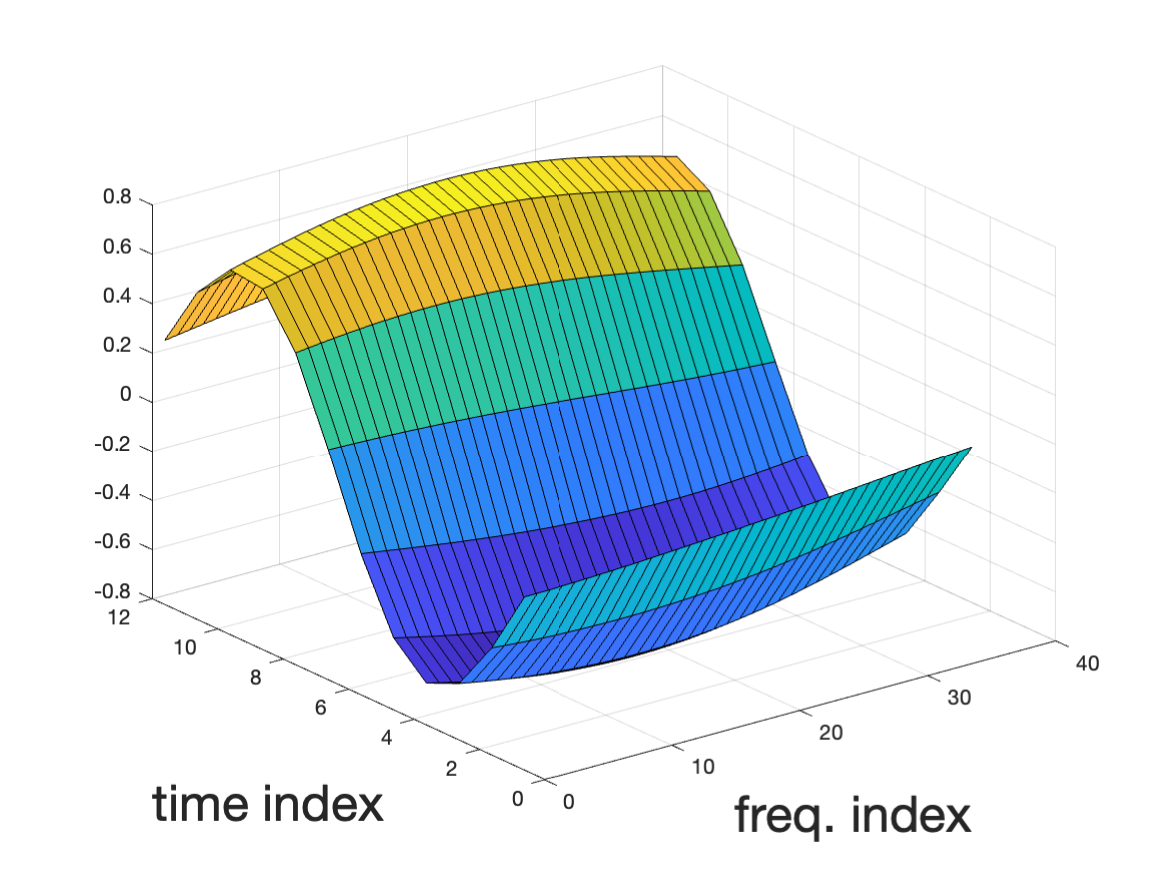}
		\caption{first basis, full block}
	\end{subfigure}
	\hfill
	\begin{subfigure}[b]{0.32\textwidth}
		\centering
		\includegraphics[width=\textwidth]{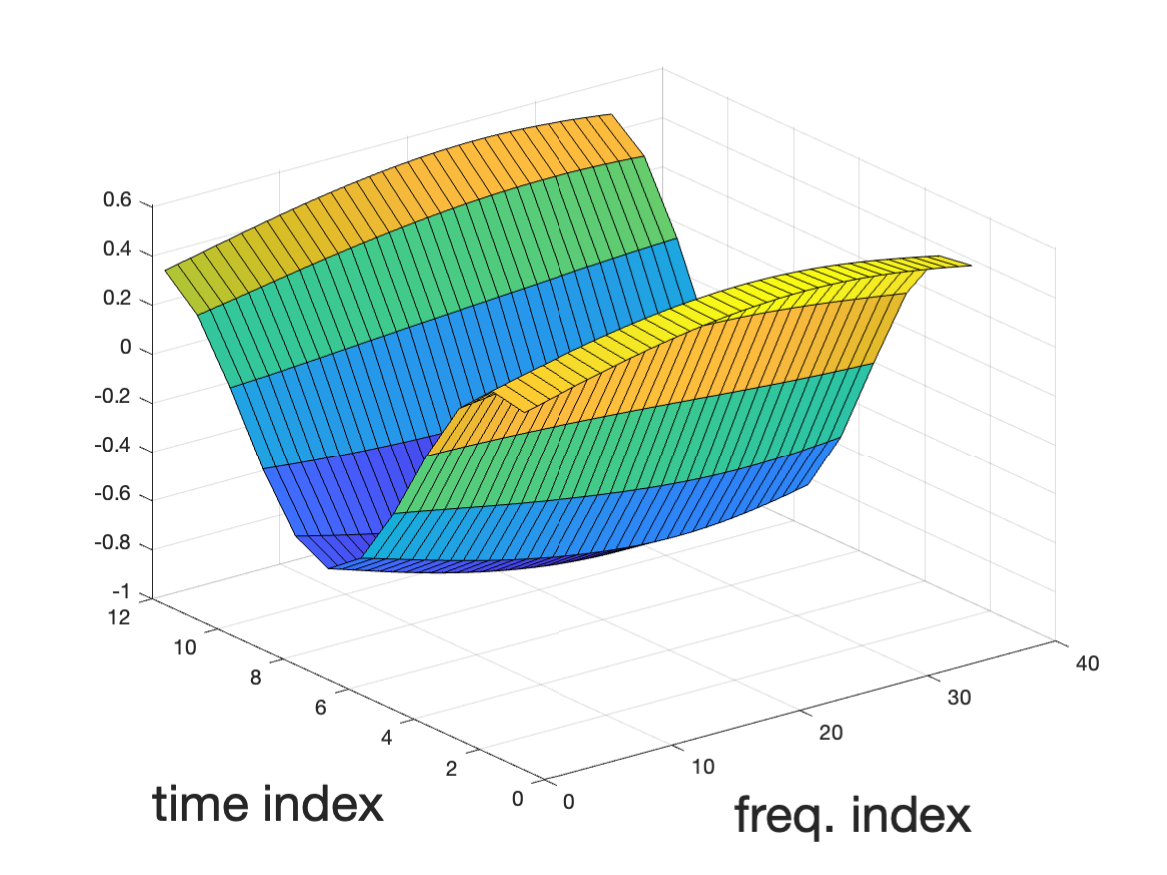}
		\caption{second basis, full block}
	\end{subfigure}
	\hfill
	\begin{subfigure}[b]{0.32\textwidth}
		\centering
		\includegraphics[width=\textwidth]{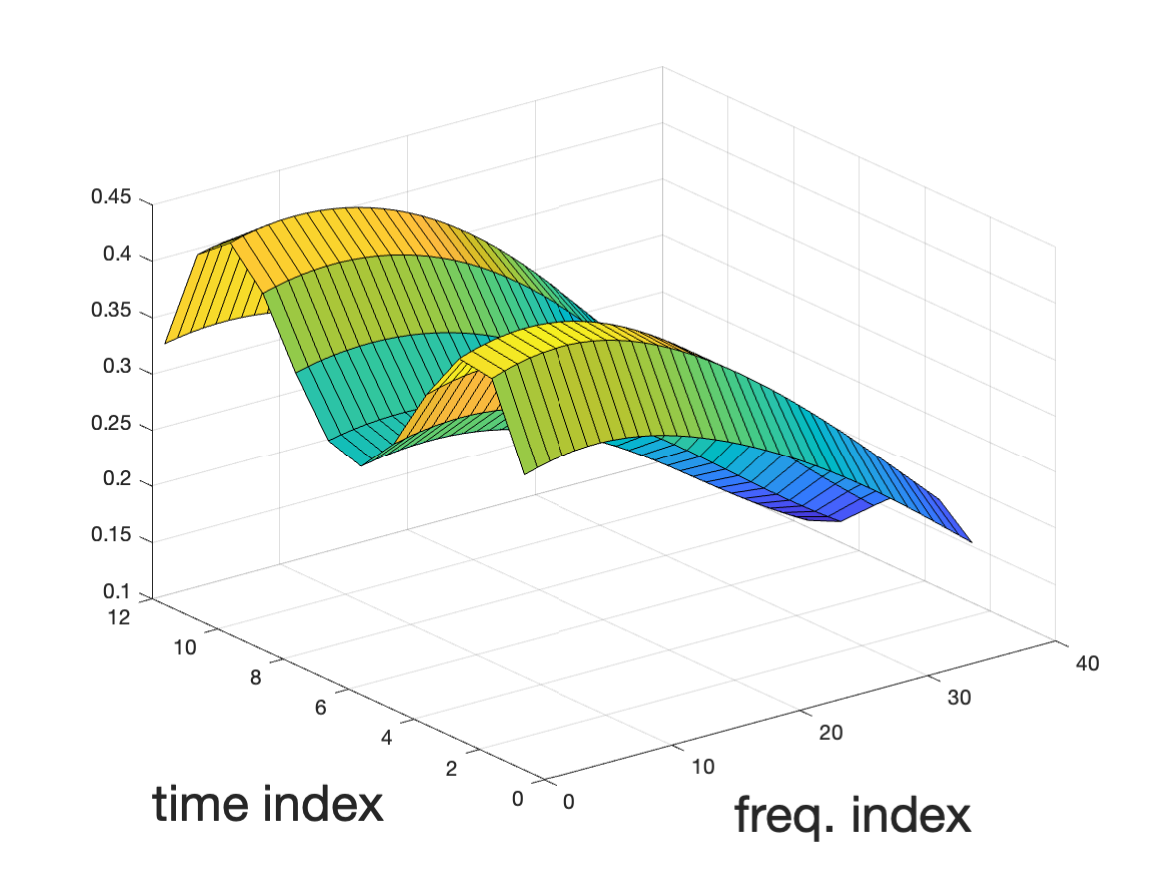}
		\caption{third basis, full block}
	\end{subfigure}
	\hfill
	\begin{subfigure}[b]{0.32\textwidth}
		\centering
		\includegraphics[width=\textwidth]{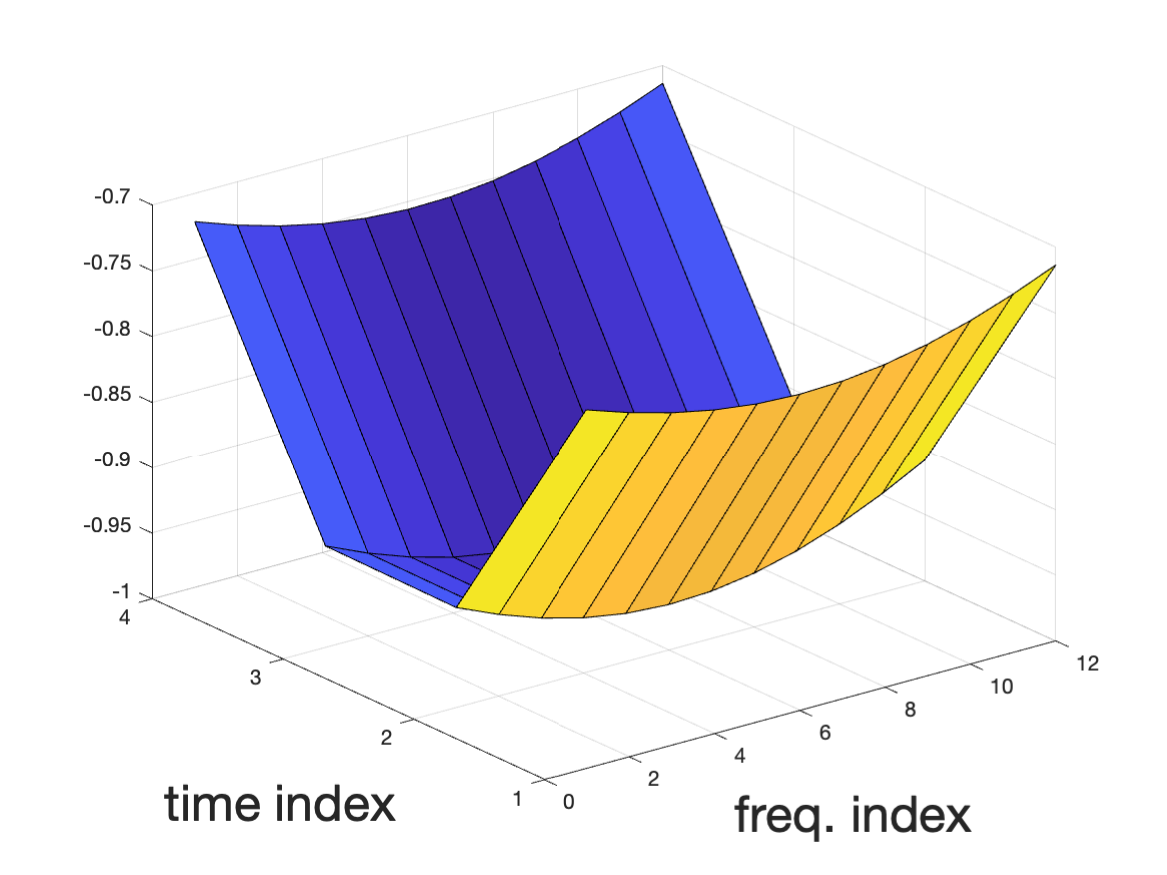}
		\caption{first basis, per sub-block}
	\end{subfigure}
	\hfill
	\begin{subfigure}[b]{0.32\textwidth}
		\centering
		\includegraphics[width=\textwidth]{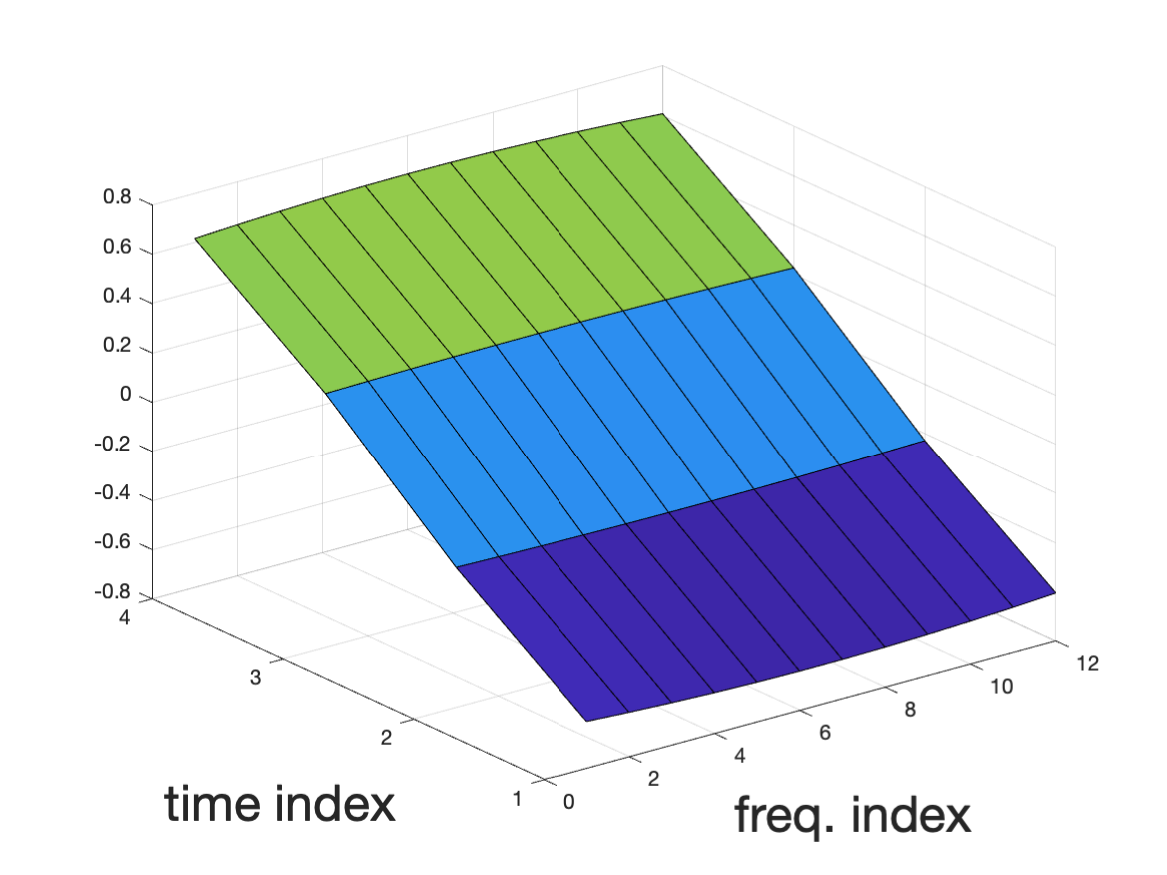}
		\caption{second basis, per sub-block}
	\end{subfigure}
	\hfill
	\begin{subfigure}[b]{0.32\textwidth}
		\centering
		\includegraphics[width=\textwidth]{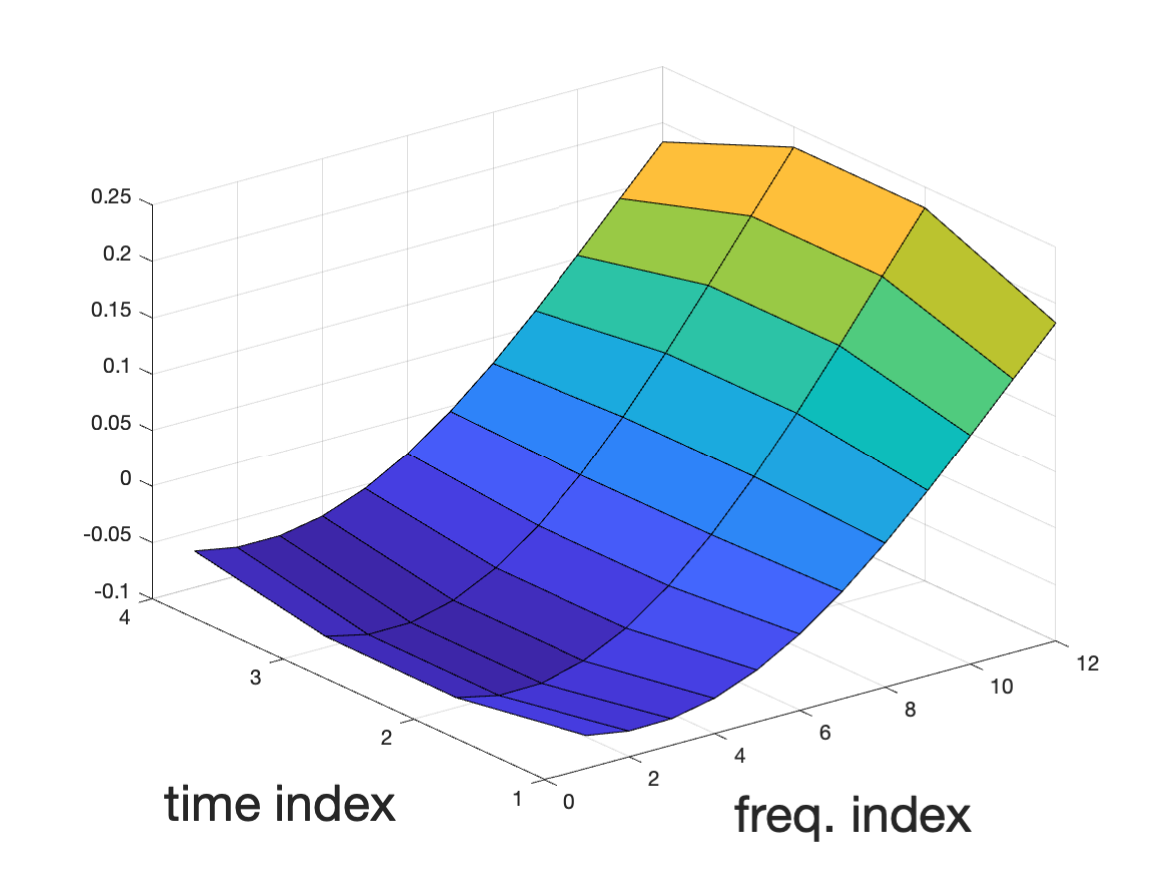}
		\caption{third basis, per sub-block}
	\end{subfigure}
	\caption{Visualization of the basis vectors.}
	\label{fig: basis-block}
%	\vspace{-0.5cm}
\end{figure*}

\begin{figure}
	\centering
	\includegraphics[width=8cm]{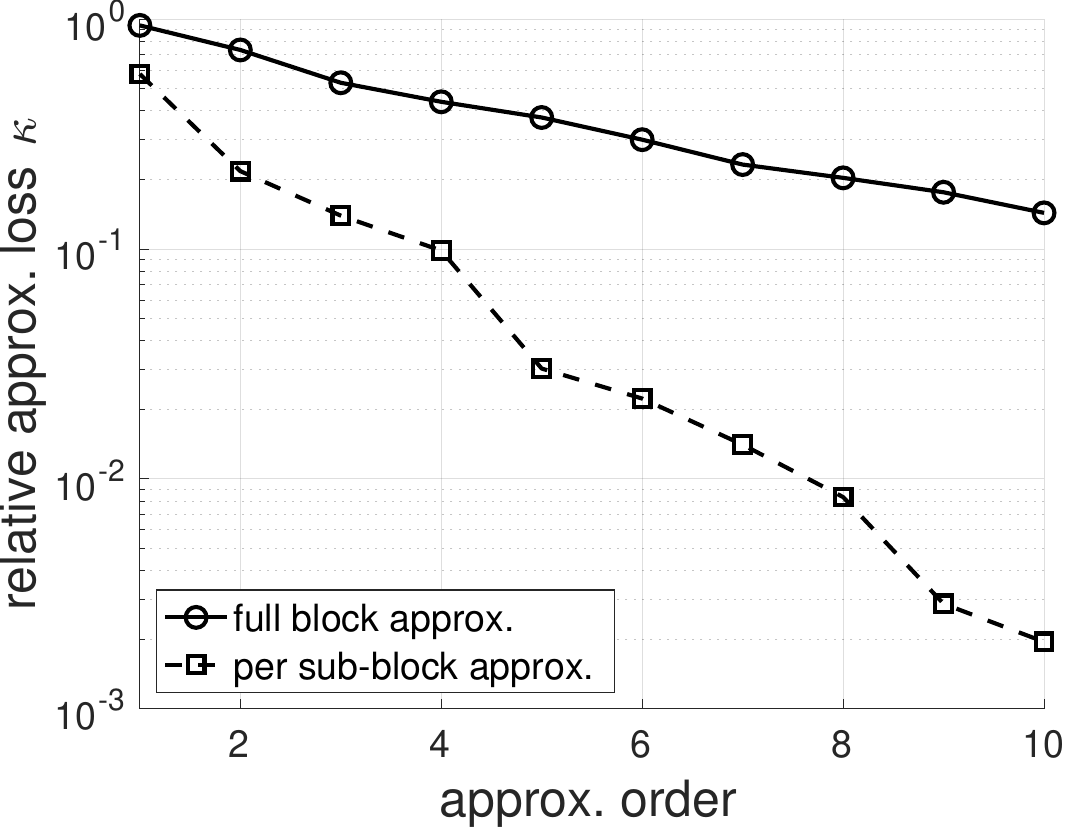}
	\caption{The value of $\kappa$ with different approximation order $N$.}
	\label{fig: kappa}
%	\vspace{-0.5cm}
\end{figure}

We choose the TDL-B channel model with a mobile speed of 120 km/h and a \gls{rms} delay spread of 1 microsecond.
We generate a sufficiently large number ($KM=10000$) of channel vectors using the procedure described in Section \ref{sec: generation of channel} to form a sample covariance matrix that is an accurate estimate of the true channel covariance matrix. 
We first check the average magnitude of the corresponding elements in the sample covariance matrix for each sub-block pair. 
As shown in Fig. \ref{fig: channel-covariance}, the diagonal elements have dominant magnitudes, suggesting strong intra-block correlations and weak inter-block correlations.
We consider two approaches to low-dimensional approximations. 
In the first approach, we directly perform the eigenvalue decomposition on the $L\times L$ sample covariance matrix to obtain the $L\times N$ basis matrix $\bG$ in \eqref{eq: channel-approx} with its columns as the dominant eigenvectors (scaled by the square roots of their corresponding eigenvalues). In the second approach, we take the average of the diagonal blocks in the sample covariance matrix corresponding to the channel sub-blocks. This new matrix has dimension $(L/P)\times(L/P)$, and we perform the eigenvalue decomposition on it to obtain a $(L/P)\times N$ basis matrix for approximating the channel variations in the sub-blocks.

One instance of the true channel (more specifically, its real part), its brute-forth full-block approximation, and the per-sub-block approximation are visualized in Fig. \ref{fig: channel-blcok}. 
The bases are visualized in Fig. \ref{fig: basis-block}. 
One can observe that by using the per-sub-block approximation, we obtain much simpler bases and a more accurate approximation. 
To quantify the approximation accuracy, we define $\mat{H} = [\cdots,\vect{h}_{km},\cdots]$ as an $L\times KM$ matrix consisting of the fading coefficients of all user-antenna pairs, $\widehat{\mat{H}} = [\cdots,\widehat{\vect{h}}_{km},\cdots]$ as the channels obtained by the low-dimensional approximation, and $\overline{\mat{H}} = [\cdots,\overline{\vect{h}}_{km},\cdots]$ with $\overline{\vect{h}}_{km} = (\frac{1}{L}\sum_{l\in[L]}h_{lkm})\vect{1}$ as the block-fading approximation. We define the metric
$
	\kappa = {\|\widehat{\mat{H}} - \mat{H} \|_\sfF}/{\|\overline{\mat{H}} - \mat{H} \|_\sfF}
$
as the relative approximation error compared with the block-fading model. The value of $\kappa$ for different approximation order $N$ is depicted in Fig. \ref{fig: kappa}. The per-sub-block approximation achieves a much more accurate approximation.

\section{Conclusion}

In this paper, we introduce a unified framework for robust user activity detection for massive access. 
Instead of assuming a block-fading channel as in most existing work, our framework allows for symbol-by-symbol variations of the channel by exploring a low-dimensional representation of the variations. 
This low-dimensional structure can be learned directly from the received pilot signals, and it provides considerable performance improvement compared with several existing baselines.
Another important component in our framework is pilot hopping which allows users to explore extra time- and/or frequency diversity.
Through case studies, we show that pilot hopping can improve the robustness of activity detection in the presence of flashlight interference or accidental blocking effects. 
However, we also observe that when those effects are not present, partitioning users into disjoint groups with dedicated radio resources achieves significantly better performance -- one should not trade sparsity for diversity in this case.
The choice of hopping patterns is also critical -- carefully designed patterns work considerably better than randomly generated ones.

\section*{Appendix: WLRMA for Covariance Estimation}

\label{sec: cov estimation}

We consider the estimation of $\bR=\bG\bG^\herm$ when an activity estimate $\widehat{\bgamma}$ is given, without restriction to the use of all-one pilots.
We note that this approach is not used in the proposed activity detection framework, as explained earlier.
We include this section mainly for completeness and to describe potential directions for future research.

For a weight matrix $\bC$, we consider the \gls{wlrma} problem
\begin{equation}
\label{eq: wlrma}
\begin{aligned}
	\min_{\bX\in\C^{L\times N}} J(\bX;\bC) \defeq& \| \bC \odot ( \bUpsilon - \bX\bX^\herm) \|_\sfF^2\\
	=& \sum_{i,j\in[L]} \left|[\bC]_{i,j}\right|^2  \left|[\bUpsilon - \bX\bX^\herm]_{i,j} \right|^2,
\end{aligned}
\end{equation}
\begin{flalign}
\label{eq: Upsilon}
	\mbox{with} && \bUpsilon \defeq (\widehat{\bSigma} - \sigma^2\bI) \oslash \bC(\widehat{\bgamma}), &&
\end{flalign}
where $\oslash$ denotes the Hadamard (element-wise) division.
One can observe from \eqref{eq: wlrma} that the problem depends only on the magnitude of the entries in $\bC$. Therefore, it is sufficient to consider only real-valued matrices $\{\bC\}$ with non-negative entries.  
When the weight matrix $\bC$ is constructed by assigning each entry the absolute value of the corresponding entry of $\bC(\widehat{\bgamma})$ in \eqref{eq: C matrix}, denoted $\bC = \bC(\widehat{\bgamma})_{\textup{abs}}$, the solution of \eqref{eq: wlrma} gives the desired estimate of the basis matrix $\bG$ in \eqref{eq: channel-approx}, i.e., $\widehat{\bG} = \argmin J(\bX;\bC(\widehat{\bgamma}))$.
Notice also that due to the Hadamard division in \eqref{eq: Upsilon}, when $\bC(\widehat{\bgamma})$ has entries with small magnitudes, the noise in $\widehat{\bSigma}$ will be magnified, and the objective becomes sensitive to the estimation error in $\widehat{\bgamma}$.
(Strictly speaking, the Hadamard division is undefined when $\bC(\widehat{\bgamma})$ has zero entries. But this case happens with zero probability when the pilots are randomly drawn from a continuous distribution.)
This partially explains the ill-conditioning of the joint estimation problem.

\subsubsection{EM Algorithm}

An \gls{em} procedure for the \gls{wlrma} problem was developed in \cite{srebro2003weighted}. 
The algorithm performs the following update in each iteration:
\begin{equation}
\label{eq: EM procedure}
	\bX_{i+1} \leftarrow \operatorname{LRA}_N(\widetilde{\bC}\odot\bUpsilon + (1-\widetilde{\bC})\odot(\bX_i\bX_i^\herm)),
\end{equation}
\begin{flalign}
\label{eq: LRA}
    \mbox{where } && \operatorname{LRA}_N(\bA) \defeq \argmin_{\bX\in \C^{L\times N}} \norm{\bA - \bX\bX^\herm}_\sfF, && 
\end{flalign}
and $\widetilde{\bC}$ is obtained by scaling $\bC$ so that the maximum element equals one, i.e., $\widetilde{\bC} \defeq \bC / \max\{[\bC]_{i,j}\}$.
Notice that the \gls{em} procedure in \eqref{eq: EM procedure} does not necessarily find the optimal solution due to the non-convexity of \eqref{eq: wlrma}.

\subsubsection{Sequential Approximation}

\begin{algorithm}[t]
	\caption{Sequential Approximation for \gls{wlrma}}
	\begin{algorithmic}[1]
		\label{alg: sequential approximation}
		\REQUIRE sample covariance $\widehat{\bSigma}$, and $\{\bC_i\}_{i=1}^I$ in \eqref{eq: weight matrices} 
		\INITIALIZE $\bUpsilon \gets (\widehat{\bSigma}-\sigma^2\bI) \oslash \bC(\widehat{\bgamma})$ \\$\bX_0 \gets \operatorname{LRA}_N(\bUpsilon)$ with $\operatorname{LRA}_N(\cdot)$ defined in \eqref{eq: LRA}
		\FOR{$i=0,1,\cdots I-1$}
		\STATE Obtain $\Delta_i$ by solving the QCQP in \eqref{eq: QCQP} 
		\STATE $\bX_{i+1} \gets \bX_i + \Delta_i$
		\ENDFOR
		\ENSURE basis matrix $\bG = \bX_I$
	\end{algorithmic}
\end{algorithm}

As it is difficult to directly solve \eqref{eq: wlrma} for $\bC = \bC(\widehat{\bgamma})_{\textup{abs}}$, the authors of \cite{lu2003new} proposed to solve \eqref{eq: wlrma} sequentially for a series of weight matrices $\bC_0,\cdots,\bC_I$. 
The first weight matrix is chosen to be $\bC_0=\bone\bone^\transp$ and the corresponding optimal solution $\bX_0$ can be easily obtained as $\operatorname{LRA}_N(\bUpsilon)$.
(When a good estimate $\widehat{\bG}_{\textup{old}}$ has been obtained from previous transmissions and if the channel covariance does not change abruptly, one may also use the corresponding $\bC(\widehat{\bgamma}_{\textup{old}})$ as $\bC_0$ and use the corresponding estimate as the initial solution.)
The final weight matrix is chosen as $\bC_I = \bC(\widehat{\bgamma})_{\textup{abs}}$ so that its optimal solution $\bX_I$ is the desired basis matrix.
The remaining weight matrices $\{\bC_i \}_{i=1}^{I-1}$ are selected as convex combinations of $\bC_0$ and $\bC_I$, i.e.,
\begin{equation}
\label{eq: weight matrices}
	\bC_i = \frac{I-i}{I}\bC_1 + \frac{i}{I}\bC_I,\quad 1\leq i \leq I-1.
\end{equation}
The key idea in \cite{lu2003new} is that the minimizer of $J(\bX;\bC)$ is a continuous function of $\bC$.
And the difference $\norm{\bX_{i+1} - \bX_i}$ can be made arbitrarily small if $\norm{\bC_{i+1}-\bC_i}$ is sufficiently small. 
This can be guaranteed when $I$ is sufficiently large.

By substituting $\bX_{i+1} = \bX_i + \Delta_i$, the \gls{wlrma} problem for the weight matrix $\bC_{i+1}$ can be reformulated as
\begin{equation}
\label{eq: wlrma-2}
\begin{aligned}
	\min_{\Delta_i\in\C^{L\times N}}  \| \bC_{i+1} \odot ( \widetilde{\bUpsilon}_i - \bX_i\Delta_i^\herm - \Delta_i \bX_i^\herm - \Delta_i\Delta_i^\herm) \|_\sfF^2
\end{aligned}
\end{equation}
with $\widetilde{\bUpsilon}_i \defeq \bUpsilon - \bX_i\bX_i^\herm$.
As discussed, when $I$ is sufficiently large, $\Delta_i$ represents a small perturbation, we can omit the second-order term $\Delta_i\Delta_i^\herm$ and add the constraint\footnote{In \cite{lu2003new}, the perturbation is measured in terms of the maximum magnitude of the entries in $\Delta_i$. However, we found that the problem can be solved more efficiently using the Frobenius norm constraint.} $\norm{\Delta_i}_\sfF \leq \varepsilon$, where $\varepsilon > 0$ is a pre-determined maximum perturbation, to obtain the approximated problem
\begin{equation}
\label{eq: QCQP}
\begin{aligned}
	\min_{\Delta_i\in\C^{L\times N}}~~&  \| \bC_{i+1} \odot ( \widetilde{\bUpsilon}_i - \bX_i\Delta_i^\herm - \Delta_i \bX_i^\herm ) \|_\sfF^2\\
	\operatorname{s.t.}~~~~& \|\Delta_i\|_\sfF \leq \varepsilon.
\end{aligned}
\end{equation}
Problem \eqref{eq: QCQP} is a convex \gls{qcqp} \gls{wrt} $\vecop(\Delta_i)$ that can be solved using standard optimization toolboxes \cite{mosek,yalmip}.

This approach is summarized in Algorithm \ref{alg: sequential approximation}.

\subsection{Numerical Results}
\label{sec: simulation cov estimation}

\begin{figure}
	\centering
	\includegraphics[width=\linewidth]{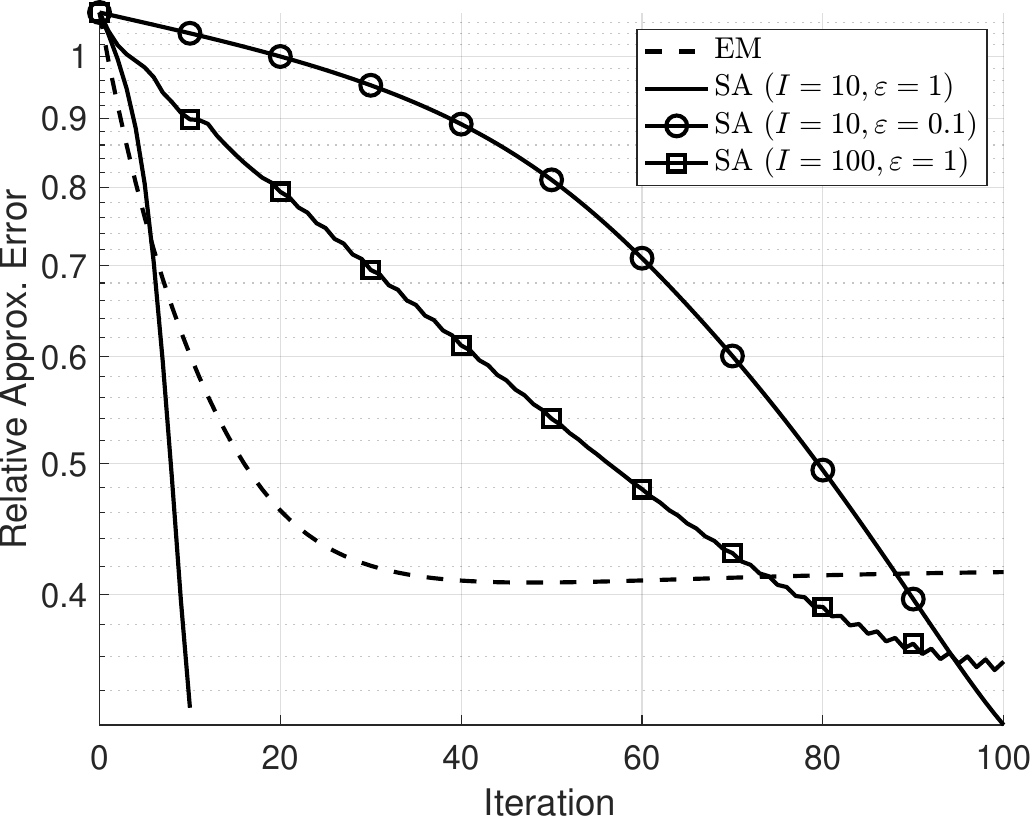}
	\caption{Convergence of the covariance estimation algorithms.}
	\label{fig: cov_est_convergence}
\end{figure}

\begin{figure}
	\centering
	\includegraphics[width=\linewidth]{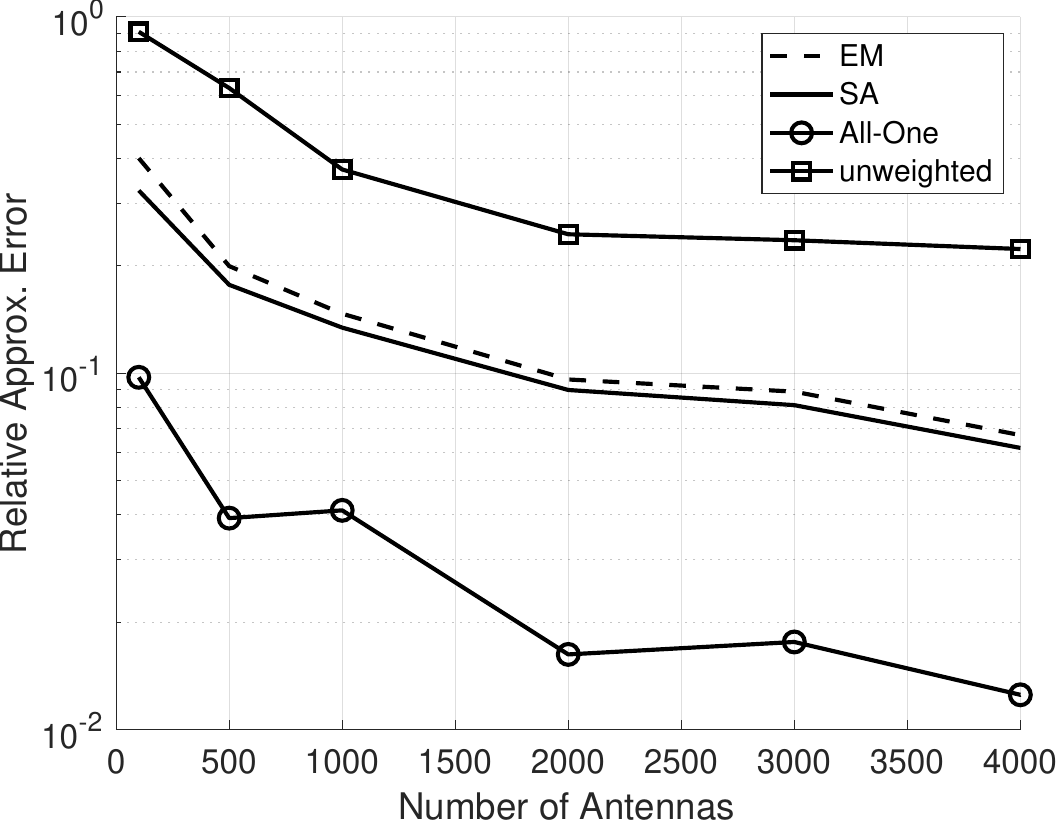}
	\caption{Comparison of the covariance estimation methods.}
	\label{fig: cov_est_comparison}
\end{figure}

We consider that 50 active users are transmitting their randomly generated Gaussian pilot sequences in a time-frequency block of size 5 by 10.
The identities of those users are assumed to be known, so that $\bC(\bgamma)$ is perfectly known.
(This is the ideal case for applying the \gls{em} and the \gls{sa} algorithms. 
When the user activities are not perfectly known, the estimation error of $\bgamma$ can further degrade the performance of channel covariance estimation. In contrast, using the all-one pilots does not require knowing the user activities.)
The performance is evaluated in terms of the relative approximation error, defined as $\norm{\bR-\widehat{\bR}}_\sfF/\norm{\bR}_\sfF$.
We first check the convergence of the covariance estimation algorithms.
As shown in Fig. \ref{fig: cov_est_comparison}, the \gls{sa} approach achieves slightly better performance than the \gls{em} algorithm.
Additionally, the \gls{sa} approach requires properly choosing the parameters $I$ and $\varepsilon$, which represent the number of sub-problems to solve and the maximum perturbation in each sub-problem, respectively. 

We then compare the performance of the \gls{em} algorithm and the \gls{sa} approach with directly solving the unweighted problem (brute-forcely using $\bC=\bone\bone^\transp$ in \eqref{eq: wlrma}), and with the case of using all-one pilots.
Suggested by the results in Fig. \ref{fig: cov_est_convergence}, we select $I=10$ and $\varepsilon=1$ for the \gls{sa} approach, and run the \gls{em} algorithm for 40 iterations.
As shown in Fig. \ref{fig: cov_est_comparison}, using the dedicated all-one pilots achieves much more accurate channel covariance estimation.
When $\bC(\bgamma)$ is diagonally dominant, the \gls{wlrma} problem becomes very sensitive to the noise in the sample covariance matrix. In this case, the \gls{em} algorithm and the \gls{sa} approach require an unrealistically large number of antennas to achieve a satisfactory performance.

Regarding the runtimes of different methods, the \gls{em} algorithm takes around 0.025 seconds, the \gls{sa} approach takes around 1.5 seconds, and using the all-one pilots takes only around 0.002 seconds.
(The \gls{qcqp} in \eqref{eq: QCQP} is solved using the YAMIP toolbox \cite{yalmip} with the MOSEK solver \cite{mosek}.)

The idea of joint activity detection and covariance estimation outlined here bears some promise, but underperforms our main algorithm (Section \ref{sec: complete framework}). It is a possible topic for future work to consider other approaches to the joint problem.

% \bstctlcite{IEEEexample:BSTcontrol}
\bibliographystyle{IEEEtran}
\bibliography{ref}

\end{document}